\documentclass{article}

\usepackage[english]{babel}

\usepackage[letterpaper,top=2cm,bottom=2cm,left=3cm,right=3cm,marginparwidth=1.75cm]{geometry}

\usepackage{amsmath}
\usepackage{graphicx}
\usepackage[colorlinks=true,allcolors=blue]{hyperref}
\usepackage{amssymb}
\usepackage{amsmath}
\usepackage{float}
\usepackage{caption}
\usepackage{rotating}
\usepackage{caption}
\usepackage{bm}
\usepackage{graphicx}
\usepackage{listings}
\usepackage[title]{appendix}
\usepackage{xcolor}

\definecolor{codegreen}{rgb}{0,0.6,0}
\definecolor{codegray}{rgb}{0.5,0.5,0.5}
\definecolor{codepurple}{rgb}{0.58,0,0.82}
\definecolor{backcolour}{rgb}{0.95,0.95,0.92}

\lstdefinestyle{mystyle}{
    backgroundcolor=\color{backcolour},   
    commentstyle=\color{codegreen},
    keywordstyle=\color{magenta},
    numberstyle=\tiny\color{codegray},
    stringstyle=\color{codepurple},
    basicstyle=\ttfamily\footnotesize,
    breakatwhitespace=false,         
    breaklines=true,                 
    captionpos=b,                    
    keepspaces=true,                 
    numbers=left,                    
    numbersep=5pt,                  
    showspaces=false,                
    showstringspaces=false,
    showtabs=false,                  
    tabsize=2
}

\title{Fast Barrier Option Pricing by the COS BEM Method in Heston Model}

\author{A. Aimi$^{1,+}$, C. Guardasoni$^{1,+,}$\footnote{Corresponding author: chiara.guardasoni@unipr.it}, L. Ortiz-Gracia$^2$, S. Sanfelici$^{3,+}$}

\date{\small{
    $^1$Dept. of Mathematical, Physical and Computer Sciences,University of Parma, Italy\\
    $^2$Dept. of Econometrics, Statistics and Applied Economics, University of Barcelona, Spain\\
    $^3$Dept. of Economics and Management, University of Parma, Italy\\
    $^+$Members of the INdAM-GNCS Research Group, Italy\vspace{0.4cm}\\
    \today
}}

\begin{document}
\maketitle

\begin{abstract}
In this work, the Fourier-cosine series (COS) method has been combined with the Boundary Element Method (BEM) for a fast evaluation of barrier option prices. After a description of its use in the Black and Scholes (BS) model, the focus of the paper is on the application of the proposed methodology to the barrier option evaluation in the Heston model, where its contribution is fundamental to improve computational efficiency and to make BEM appealing among Finance practitioners as a valid alternative to Monte
Carlo (MC) or other more traditional approaches. An error analysis is provided on the number of terms used in the Fourier-cosine series expansion, where the error bound estimation is based on the characteristic function of the log-asset price process.
\end{abstract}



\pagestyle{myheadings}
\thispagestyle{plain}
\markboth{}{SABO+cos method}

\section{Introduction}
\label{sec;introduction}
A European option is a financial derivative contract that gives the buyer the right to buy (\emph{call option}) or sell (\emph{put option}) a
particular asset at a fixed maturity or expiry $T$ and at a predetermined exercise or strike price $K$. In
the case of a barrier option, this right is activated (knock-in) or extinguished (knock-out) when the
underlying asset reaches a certain barrier price during the time interval $[0, T]$. Barrier options were
created to provide the hedge of an option at a lower premium than a conventional option and are traded
in large volumes.
There are four main types of barrier options
that can have either call (buy) or put (sell) feature: down-and-in, down-and-out, up-and-in and
up-and-out. The “down” and “up” terms refer to the position of the barrier in relation
to the initial underlying price. Despite being frequently traded nowadays,
barrier options are still known as exotic options since they cannot be replicated by
a finite combination of standard products, i.e., vanilla call and put options, future
contracts, etc., and a closed-form pricing formula is not available unless we assume a Black-Scholes (BS) framework for the underlying asset.

Traditional numerical approaches to barrier option pricing are based on lattice models \cite{BoyleLau1994} and Monte Carlo (MC) or conditional sampling MC methods \cite{GlassermanStaum2001} that are affected by high computational costs and inaccuracy due to their intrinsic slow convergence.
Other possible approaches are offered by the finite difference method (FDM), the finite element method (FEM) and finite volume methods (see among others \cite{GolbabaiBallestraAhmadian2014,VetzalForsythZvan2000,Ritchken1991,RogersZane1997}). 
In \cite{Chiarella2012}, the authors develop a method of lines approach to evaluate the option price, which is able to efficiently handle
both continuously monitored and discretely monitored barrier options and also early exercise features.
Finally, the paper by \cite{LiptonMcGhee2002} proposes a formalism for pricing and hedging exotic options on forex rates under various model frameworks (including, as special cases, local, jump-diffusion and stochastic volatility models).

More recently, several significant contributions have been proposed to develop efficient numerical pricing algorithms for exotic options in models beyond the BS setting.
Focusing on the specific field of barrier options, we cite the quasi-MC method of \cite{AchtsisCoolsNuyens2013}, which introduces a conditional sampling method to deal with barriers combined with a path construction method, and the Heath–Platen estimator of \cite{CoskunKorn2018}, which uses a Generalized Black-Scholes (GBS) process as the basis for an expansion around the GBS-option pricing formula. 
Other related approaches derive semi-analytical formulas for the prices of path-dependent options by conditioning with respect to the variance path and obtaining the joint probability distribution between the logarithmic spot price and its maximum/minimum. The authors in \cite{LiptonSepp2022} combine the one-dimensional Monte Carlo simulation and the semi-analytical one-dimensional heat potential method to design an efficient technique for pricing barrier options on assets with correlated stochastic volatility. They first condition the price dynamics on a given volatility path and apply the method of heat potentials to solve the conditional problem in closed form. Then, they run an outer loop by generating volatility paths via the Monte Carlo method and averaging over the space of variance trajectories. 

Another stream of research applies the so-called Fourier-cosine series (COS) method by \cite{FangOosterlee2008} to discretely monitored barrier and Bermudan options under the Heston and L\'evy models. See, for instance, \cite{FangOosterlee2009,FangOosterlee2011,Borovykh2017,Borovykh2018}.

A recent field of research based on wavelets offers efficient solutions in the context of option pricing. The paper by \cite{MareeOrtiz-GraciaOosterlee2017} lies within the class of pricing methods based on the expected discounted payoff pricing formula for discretely monitored barrier options under exponential L\'evy dynamics. The authors present a valuation method where the Fourier inversion step is carried out by approximating the unknown density function by means of a finite combination of wavelets basis functions. This pricing machinery was inspired in previous works initially developed by \cite{Ortiz-GraciaOosterlee2013,Ortiz-GraciaOosterlee2016} for pricing European options where the asset price process is governed by L\'evy and Heston models. The basis functions used in \cite{Ortiz-GraciaOosterlee2013} are the Haar wavelets, while Shannon wavelets are employed in \cite{Ortiz-GraciaOosterlee2016,MareeOrtiz-GraciaOosterlee2017} giving birth to the so-called SWIFT method. The compact support feature of the Haar family allows for the efficient pricing of options with very long maturities, where the COS method \cite{FangOosterlee2008} fails, although the COS method is preferred for normal maturities due to the regularity of the density functions that we typically encounter in option pricing. Shannon wavelets are compactly supported in the Fourier domain, and they are regular functions in the time domain. The band-limited characteristic of the Shannon family allows for a robust valuation of the option, since we have an a priori knowledge of the parameters associated to the numerical method (see \cite{MareeOrtiz-GraciaOosterlee2017} for the details). Despite its robustness, SWIFT method is more time consuming than COS method, and that is the reason why we select the COS method in the present work to carry out the Fourier inversion step.

The present paper builds upon the work by \cite{GuardasoniSanfelici2016}, where a semi-analytical resolution method has been applied to the Heston model for pricing barrier options, taking advantage of the BEM features: the discretization is applied only to the boundary of the model problem domain, represented in this framework by the barriers; the solution in the interior of the domain is approximated with a rather high convergence rate and can be evaluated at any specific point
of the domain (e.g. the current asset price), as required during financial transactions,
avoiding its computation everywhere on a defined grid.
The BEM method is based on an integral representation of the starting differential problem solution. For plain vanilla options, this integral formulation
reduces to the risk-neutral evaluation formula of option pricing, i.e., the discounted expectation of the final payoff under suitable
probability measure (see \cite{Duffie1996}). For barrier options, in a stochastic volatility framework as configured in Heston model \cite{Heston1993}, the integral representation formula of the option price depends on time, volatility and the underlying asset value but, using the
boundary condition at the barrier, we obtain a Fredholm integral equation of the first kind that can be solved
by discretization in time and variance only. Therefore, the dimensionality of the problem is reduced by one compared to other domain methods such as FEM or FDM, avoiding the discretization of the underlying asset space. 
The integration domain is bounded in time and unbounded in
variance; however, exploiting the far-field properties of the option price and of the
kernels appearing in the integrals, it can be truncated to apply standard quadrature rules and in particular we simply apply a Matlab adaptive quadrature function.

The main drawback of this method concerns
the necessary knowledge of the fundamental solution for the differential equation
governing the price of the contingent claim, namely the conditional (i.e., transition)
probability density function (PDF) of the underlying asset price process, that is generally explicitly available only for certain partial differential equations (PDEs). For more general PDEs, such as those related to stochastic volatility models, 
it may be available only through its Fourier transform, i.e. the characteristic function.
In particular, when pricing exotic options in the Heston framework, the joint transition PDF of stock log-price
and variance must be computed. When this is the case, the computational bottleneck of the BEM approach, and integral formulations in general, reduces to the numerical evaluation of complex integrals.
Quadrature rule based techniques are not efficient when computing Fourier transform integrals, due to the fact that, as the integrands are highly oscillatory, a relatively fine grid has to be used to get the desired accuracy.
In this paper we apply the COS method, based on Fourier-cosine expansions, in the context of numerical
integration. This method proved to offer a highly efficient way to recover the density from the characteristic function.
We show here that it can further improve the speed of the BEM for pricing barrier options in the Heston framework, where
we do not know explicitly the analytical expression for the fundamental solution and a Fourier inversion is required. Further, 
we provide an error analysis on the number of terms used in the Fourier-cosine series expansion. The estimated error bound is based on the characteristic function, which is known for most of the interesting processes in finance, rather than on the theoretical properties of the PDF of the log-asset price, which is in general unknown. To the best of our knowledge, this is the first time that this issue is addressed.

The paper is organized as follows. In Section 2, we summarize the COS method and provide an error analysis. In Section 3
we illustrate the BEM approach combined with the COS method on 
the simple BS model with time-dependent risk-free interest rate. In Section 4, the methodology is extended to the Heston stochastic
volatility model. After recalling the main issues concerning the option
pricing under the Heston framework, we provide an integral representation formula which
allows us to price barrier options starting from a computation of the joint transition
PDF of stock log-price and variance by Fourier inversion using the COS method. 
A wide variety of numerical experiments and results validating the efficiency of the proposed approach is given throughout the paper. Finally, conclusions are provided in Section 5.

\section{The COS method}\label{sec;COS}
The COS method belongs to the class of Fourier inversion methods, and it is used to recover the PDF of the log-asset price process at terminal time $T$ from its characteristic function, which is, as mentioned above, the Fourier transform of the PDF. 

Generally speaking, a PDF $G(x)$ and its characteristic function $\mathcal{G}(\omega)$ form a Fourier pair, i.e.
\begin{eqnarray}
\mathcal{G}(\omega)=\mathcal{F}_x[G(x)](\omega):=\int_{\mathbb{R}}e^{i\omega x}G(x)dx, 
\\ \quad G(x)=\mathcal{F}_\omega^{-1}[\mathcal{G}(\omega)](x):=\frac{1}{2\pi}\int_{\mathbb{R}}e^{-ix \omega}\mathcal{G}(\omega)d\omega.
\label{densidad}
\end{eqnarray}

For a function supported on $[0,\pi]$, the cosine expansion reads
\begin{equation}
G(\theta)=\sum_{n=0}^{\infty}{}^{\prime}A_{n}\cos(n\theta), \quad \text{with} \quad A_{n}=\frac{2}{\pi}\int_{0}^{\pi}G(\theta)\cos(n\theta)d\theta\label{cosineexp},
\end{equation}
where $\sum^{\prime}$ indicates that the first term in the summation is weighted
by one-half. For functions supported in any other finite interval,
say $[a,b] \subset \mathbb{R}$, the Fourier-cosine series expansion
can easily be obtained via the following change of variables
\begin{equation}
\theta:=\frac{x-a}{b-a}\pi, \quad x=a+\frac{b-a}{\pi}\theta.\label{aux}
\end{equation}
It then reads
\begin{equation}\label{fouriercosserexp}
G(x)=\sum\limits _{n=0}^{\infty}{ }^{\prime}A_{n}\cos\left(n\pi\frac{x-a}{b-a}\right), \quad \text{with} \quad A_{n}=\frac{2}{b-a}\int_{a}^{b}G(x)\cos\left(n\pi\frac{x-a}{b-a}\right)dx.
\end{equation}
Then, $A_{n}$ is approximated\footnote{In fact, by the definition of characteristic function, we get
$$\begin{array}{rcl}
A_{n}&=&\displaystyle\frac{2}{b-a}\int_{a}^{b}G(x)\cos\left(n\pi\frac{x-a}{b-a}\right)dx=\frac{2}{b-a}\textrm{Re}\left\{ \int_a^b G(x)\left(\cos\left(\frac{n\pi(x-a)}{b-a}\right)+i\sin\left(\frac{n\pi(x-a)}{b-a}\right)\right)dx\right\}\\
&=&\displaystyle\frac{2}{b-a}\textrm{Re}\left\{ \int_a^b G(x)\exp\left(i\frac{n\pi(x-a)}{b-a}\right)dx\right\}=\frac{2}{b-a}\textrm{Re}\left\{\exp\left(-i\frac{n\pi a}{b-a}\right) \int_a^b G(x)\exp\left(i\frac{n\pi x}{b-a}\right)dx\right\}\approx F_{n}. \label{Fk}
\end{array}$$} by
\begin{equation}
\displaystyle  F_{n}:=\frac{2}{b-a}\textrm{Re}\left\{\exp\left(-i\frac{na\pi}{b-a}\right) \mathcal{G}\left(\frac{n\pi}{b-a}\right)\right\}, \label{Fk_0}
\end{equation}
where $\textrm{Re}(z)$ denotes the real part of $z$. Finally, we replace $A_n$ by $F_n$ in (\ref{fouriercosserexp}) and truncate the series summation such that
\begin{equation}\label{eq:density_cos}
G(x) \approx \sum_{n=0}^{N_F-1}{ }^{\prime}F_{n}\cos\left(n\pi\frac{x-a}{b-a}\right),
\end{equation}
which allows to recover the PDF from its characteristic function $\mathcal{G}(\cdot)$.
When $[a,b]$ is conveniently chosen the overall error of the COS method is dominated by the series truncation error $\epsilon_t$ of the density function (see \cite{FangOosterlee2008} for details), i.e.

\begin{equation}\label{eq:truncerror}
\epsilon_t:=\sum_{n=N_F}^{\infty} \left |A_{n}\right|.
\end{equation}

Finally, the pricing of plain vanilla options with the COS method, as proposed by \cite{FangOosterlee2008}, involves the computation of the density coefficients in expression (\ref{Fk_0}), as well as the payoff coefficients. The payoff coefficients are calculated exactly, as the integral on $[a,b]$ of the option payoff multiplied by the cosine basis.

We study the error \eqref{eq:truncerror} when the series in expression (\ref{fouriercosserexp}) is truncated to the first $N_F$ terms, and is finally approximated by the finite sum in expression (\ref{eq:density_cos}). An error analysis is given in \cite{FangOosterlee2008} based on regularity assumptions of the PDF of the log-asset price at maturity. Since the PDF is generally unknown in the option pricing problem, the selection of the truncation parameter $N_F$ is a matter of trial and error. To the best of our knowledge, this is the first time that an assessment of the series truncation error is based on the characteristic function $\mathcal{G}(\omega)$ of the log-asset price at maturity, which is known in closed form for most of the interesting models employed in finance. We approximate the truncation error in expression (\ref{eq:truncerror}) by,
\begin{equation}\label{eq:truncerrorbar}
\epsilon_t\approx\bar{\epsilon}_t:=\sum_{n=N_F}^{\infty} \left |F_{n}\right|\leq\frac{2}{b-a}\sum_{n=N_F}^{\infty}
\left|\exp\left(-i\frac{na\pi}{b-a}\right) \mathcal{G}\left(\frac{n\pi}{b-a}\right)\right|.
\end{equation}
We observe that,
\begin{equation}\label{mod}
\bar{\epsilon}_t \le \frac{2}{b-a}\sum_{n=N_F}^{\infty}\left|\mathcal{G}\left(\frac{n\pi}{b-a}\right)\right|,
\end{equation}
and we can therefore give an estimation of the error in terms of the modulus of the Fourier transform of $G$. It is worth underlining that the error bound estimation given in expression (\ref{mod}) is an approximation of $\epsilon_t$ but it is not necessarily an upper bound of $\epsilon_t$. Anyway, starting from \eqref{mod}, an indication on the selection of the parameter $N_F$ can be deduced, as shown in the following Sections \ref{sec:COSerroranalysis} and \ref{sec:COSerroranalysis_H}.

\section{BS model with time-dependent risk-free interest rate}
\label{sec;model}
In the BS model \cite{BlackScholes1973}, beside some structural parameters, the option value $V$ depends on the current time $t$ and underlying asset value $S$. Here we assume the volatility of the underlying asset $\sigma$ to be constant while the risk-free interest $\bar{r}$ deterministically dependent on time.

For simplicity of exposition, let us consider an up-and-out barrier put option with European exercise style and strike price $E$, so that the option price vanishes if the underlying asset grows up enough to breach an assigned upper barrier $S_B$ before the expiry date $T$. Then, the differential model problem to solve is
\begin{eqnarray}\label{BS0}
\displaystyle  \frac{\partial V}{\partial t}(S,t)+\frac{\sigma^2}{2} S^2\frac{\partial^2 V}{\partial S^2}(S,t)+\bar{r}(t)S\frac{\partial V}{\partial S}(S,t)-\bar{r}(t)V(S,t)=0\qquad \forall S\in (0,S_B),\, t\in[0,T)\\
\label{Payoff}
\displaystyle V(S,T)=\max(E-S,0)\hspace{7.8cm}\forall S\in(0,S_B)\\
\label{BarrierCond}
\displaystyle V(S_B,t)=0\hspace{10.05cm}\forall t\in[0,T)
\end{eqnarray}
Now, we can follow some steps of the procedure described in \cite{Guardasoni2018}.

$\bullet$ In the risk-neutral framework, performing the classical changes of variables
\begin{equation}\label{Change}
\displaystyle  V(S,t)=u(x,t)e^{-\int^T_t \bar{r}(t')dt'},\,S=e^x,\, \tau=T-t,\,\bar{r}(t)=\bar{r}(T-\tau)=:r(\tau)
\end{equation}
the resulting diffusion problem reads
\begin{eqnarray}\label{BS}
  \displaystyle\frac{\partial u}{\partial \tau}(x,\tau)-\frac{\sigma^2}{2}\frac{\partial^2 u}{\partial x^2}(x,\tau)-\left(r(\tau)-\frac{\sigma^2}{2}\right)\frac{\partial u}{\partial x}(x,\tau)=0\qquad  x\in\Omega=(-\infty,B),\,\tau\in(0,T]\vspace{0.2cm}\\
  \displaystyle u(x,0)=\max(E-e^x,0)=:u_0(x)\hspace{7.2cm} x\in\Omega\vspace{0.2cm}\\
  \displaystyle u(B,\tau)=0\hspace{7.8cm} B:=\log(S_B),\,\,\tau\in[0,T]
\end{eqnarray}

$\bullet$ Applying the Green theorem (for the proof look at \cite{Guardasoni2018}), the unknown $u(x,\tau)$ can be represented in an integral form as
\begin{equation}\label{solBarrier}
\begin{array}{l}
\displaystyle u(x,\tau)= \int_{-\infty}^{B}u_0(y)G(y,0;x,\tau)dy+\int_{0}^\tau\frac{\sigma^2}{2}G(B,s;x,\tau)\frac{\partial u}{\partial y}(B,s)ds\qquad x\in\Omega, \tau\in(0,T],
\end{array}
\end{equation}
where the transition probability density function (or fundamental solution) is known in closed form
\begin{equation}\label{FundSol}
\displaystyle G(y,s;x,\tau)=\frac{1}{\sigma\sqrt{2\pi(\tau-s)}}
\exp\left\{-\frac{\big[y-x+\frac{\sigma^2}{2}(\tau-s)-\int_s^\tau\overline{r}(v)dv\big]^2}
{2\sigma^2(\tau-s)}\right\}\,,\quad \tau>s.
\end{equation}

\noindent \textbf{Remark.} The well-known formula for pricing European options without barriers is given only by the first term in \eqref{solBarrier}
\begin{equation}\label{sol_no_Barrier}
\begin{array}{l}
\displaystyle u(x,\tau)= \int_{-\infty}^{\infty}u_0(y)G(y,0;x,\tau)dy\qquad x\in(-\infty,\infty), \tau\in(0,T]
\end{array}
\end{equation}
and, in \cite{FangOosterlee2008}, the authors suggest to introduce the COS method, that requires, instead of $G$, the knowledge of the characteristic function $\mathcal{G}$, i.e. its Fourier transform w.r.t. the asset variable $y$:	
\begin{equation}\label{char_fun}
		\mathcal{G}(\omega,s;x,\tau)=\displaystyle e^{\mathbf{i}\omega x+\mathbf{i}\omega\int_s^\tau r(v)dv-\mathbf{i}\omega\frac{\sigma^2}{2}(\tau-s)-\omega^2\frac{\sigma^2}{2}(\tau-s)}.
\end{equation}
Then
\begin{equation}\label{Fund_char}
  G(y,s;x,\tau)=\mathcal{F}_\omega^{-1}[\mathcal{G}(\omega,s;x,\tau)](y,s;x,\tau)
\end{equation}
can be approximated by its truncated Fourier-cosine expansion \begin{equation}
\sum_{n=0}^{N_F-1}{ }^{\prime}F_{n}[s;x,\tau]\cos\left(n\pi\frac{y-a}{b-a}\right)
\end{equation}
where the coefficients can be computed directly from the characteristic function
\begin{equation}
F_{n}[s;x,\tau]=\frac{2}{b-a}\textrm{Re}\left\{ \mathcal{G}\left(\frac{n\pi}{b-a},s;x,\tau\right)\exp\left(-i\frac{na\pi}{b-a}\right)\right\}, \label{Fk_BS0}
\end{equation}
so that the solution \eqref{sol_no_Barrier} is approximated by the rapidly convergent series
\begin{equation*}
u(x,\tau)\approx{\sum_{n=0}^{N_F-1}}^{\prime}F_{n}[0;x,\tau]\int_a^b u_0(y)\cos\left(n\pi\frac{y-a}{b-a}\right)dy \qquad x\in(-\infty,\infty), \tau\in(0,T]\,.
\end{equation*}
Fang and Oosterlee \cite{FangOosterlee2008} then compute the integrals in the summation from the cosine series coefficients of the payoff function $u_0(y)$ in $y$, to get their COS pricing formula.
Unfortunately, this strategy cannot be straightforwardly applied to equation  \eqref{solBarrier} because of the second term where the function $\frac{\partial u}{\partial y}(B,s)$ is still unknown.

$\bullet$ Letting $x\rightarrow B$ in \eqref{solBarrier} and applying the vanishing boundary condition at the barrier, we obtain the Boundary Integral Equation (BIE)
\begin{equation}\label{BIE}
\begin{array}{l}
\displaystyle 0=u(B,\tau)= \int_{-\infty}^{B}u_0(y)G(y,0;B,\tau)dy+\int_{0}^\tau\frac{{\sigma}^2}{2}\frac{\partial u}{\partial y}(B,s)G(B,s;B,\tau)ds\,.
\end{array}
\end{equation}
whose sole unknown is the function $\frac{\partial u}{\partial y}(B,s)$.\\ The BEM consists of two steps: first numerically solve \eqref{BIE} at the boundary getting $\frac{\partial u}{\partial y}(B,s)$ and then insert it in the representation formula \eqref{solBarrier} to recover the solution $u$ at any desired point of the whole domain. The two steps can be done by the Fourier inverse transform of the characteristic function to recover the PDF $G$ and this can be easily achieved by the truncated Fourier-cosine expansion, extracting the series coefficients directly from the characteristic function as explained above.\\ The option price $V(S,t)$, solution to the differential problem \eqref{BS0}-\eqref{BarrierCond}, is then evaluated by transforming back with formulas \eqref{Change}.

\subsection{Approximation of the BS BIE solution by the COS method}
\label{sec;NumApproxBIE}
The time interval $[0,T]$ is subdivided in $N_{\Delta t}\in\mathbb{N}^+$ intervals of length $\Delta t=T/N_{\Delta t}$, 
$$
\displaystyle t_{k}=k\Delta t,\quad k=0,\ldots,N_{\Delta t}\,.
$$
The unknown $\frac{\partial u}{\partial y}(B,s)$ is approximated in time by piecewise constant basis functions $\varphi_k(s),\,k=1,\ldots,N_{\Delta t}$:
\begin{equation}\label{approx}
\displaystyle \frac{\partial u}{\partial y}(B,s)\approx\phi(s):=\sum_{k=1}^{N_{\Delta t}}\alpha_k\varphi_k(s)
\end{equation}
and then equation \eqref{BIE} is evaluated at the collocation points $\overline{t}_j,\,j=1,\ldots,N_{\Delta t}$
$$
\begin{array}{l}
\displaystyle 0=u(B,\overline{t}_j)= \int_{-\infty}^{B}u_0(y)G(y,0;B,\overline{t}_j)dy+\int_{0}^{\overline{t}_j}\sum_{k=1}^{N_{\Delta t}}\alpha_k\varphi_k(s)\frac{\sigma^2}{2}G(B,s;B,\overline{t}_j)ds
\end{array}
$$
choosing, as collocation points, the centers of intervals $[t_{j-1},t_j]$
$$
\overline{t}_j=\frac{t_j+t_{j-1}}{2},\quad j=1,\ldots,N_{\Delta t}\,.
$$
This procedure leads to a linear system
\begin{equation}\label{system}
\mathcal{A}\alpha=\mathcal{F}
\end{equation}
whose unknown is the vector $\alpha$ of coefficients in (\ref{approx}).

Due to the properties of the fundamental solution \eqref{FundSol}, the matrix $\mathcal{A}$ has lower triangular structure
\begin{equation}\label{struct}
\mathcal{A}=\left[
              \begin{array}{ccccc}
                A_{11} & 0 & 0 & \cdots & 0 \\
                A_{21} & A_{22} & 0 & \cdots & 0 \\
                A_{31} & A_{32} & A_{33} & \cdots & 0 \\
                \vdots & \cdots & \ddots & \ddots & \vdots \\
                A_{N_{\Delta t}1} & A_{N_{\Delta t}2} & \cdots & A_{N_{\Delta t}N_{\Delta t}-1} & A_{N_{\Delta t}N_{\Delta t}} \\
              \end{array}
            \right]
\end{equation}
and the COS method has been applied for the evaluation of its entries:\\ for $j,k=1,\ldots,N_{\Delta t}$, $j\geq k$,
\begin{equation}\label{matrix}
\begin{array}{l}
\displaystyle\mathcal{A}_{jk}=\int_0^{\overline{t}_j}\varphi_k(s)
\frac{\sigma^2}{2}G(B,s;B,\overline{t}_j)ds=\int_{t_{k-1}}^{\min(t_{k},\,\bar t_j)}\frac{\sigma^2}{2}G(B,s;B,\overline{t}_j)ds\vspace{0.2cm}\\
\displaystyle=\int_{t_{k-1}}^{\min(t_{k},\,\bar t_j)}\frac{\sigma^2}{2}
\int_{-\infty}^{+\infty}\frac{e^{-\mathbf{i}\omega B}}{2\pi}
  \mathcal{G}(\omega,s;B,\overline{t}_j)d\omega ds\vspace{0.2cm}\\
\displaystyle  \approx\int_{t_{k-1}}^{\min(t_{k},\,\bar t_j)}\frac{\sigma^2}{2}
\left\{\sum_{n=0}^{N_F-1}{}^{\prime}F_n[B,s,\bar t_j]\cos\left(n\pi\frac{B-a}{b-a}\right)\right\}ds
\end{array}
\end{equation}
having replaced the fundamental solution $G$ by a cosine expansion truncated to $N_F$ terms with coefficients
\begin{equation}\label{F_k}
\begin{array}{l}
\displaystyle F_n[B,s,\bar t_j]=\frac{2}{b-a}\textrm{Re}\left\{\mathcal{G}\left(\frac{k\pi}{b-a},s;B,\overline{t}_j\right)e^{-\mathbf{i}k\pi\frac{a}{b-a}}\right\},\qquad n=0,\ldots,N_F-1,
\end{array}
\end{equation}
and bounds
\begin{equation}\label{bounds}
\begin{array}{l}
\displaystyle a=\int_s^{\bar t_j}\overline{r}(v)dv-\frac{\sigma^2}{2}\left(\overline{t}_j-s\right)-L\sqrt{\sigma^2\left(\overline{t}_j-s\right)}+B\vspace{0.2cm}\\
\displaystyle b=\int_s^{\bar t_j}\overline{r}(v)dv-\frac{\sigma^2}{2}\left(\overline{t}_j-s\right)+L\sqrt{\sigma^2\left(\overline{t}_j-s\right)}+B
\end{array}
\end{equation}
Based on the error analysis of the following Section \ref{sec:COSerroranalysis}, the parameters of the cosine expansion are set as $L=10, N_F=50$.\\

The right-hand side entries evaluated by the COS method are
\begin{equation}\label{rhs}
\begin{array}{l}
\displaystyle\mathcal{F}_j=-\int_{-\infty}^{B}u_0(y)G(y,0;B,\overline{t}_j)dy\vspace{0.2cm}\\
\displaystyle=-\int_{-\infty}^{\min(B,\log(E))}(E-e^y)
\int_{-\infty}^{+\infty}\frac{e^{-\mathbf{i}\omega y}}{2\pi}
  \mathcal{G}(\omega,0;B,\overline{t}_j)d\omega dy\vspace{0.2cm}\\
\displaystyle\approx-\int_{a}^{\min(B,\log(E),b)}(E-e^y)
\left\{\sum_{n=0}^{N_F-1}{}^{\prime}F_n[B,0,\bar t_j]\cos\left(n\pi\frac{y-a}{b-a}\right)\right\} dy\,,
\end{array}
\end{equation}
but in this case the coefficients $F_n$ are independent from the integration variable $y$ and the integration of the payoff function times $\cos\left(n\pi\frac{y-a}{b-a}\right)$ could be performed analytically as in \cite{FangOosterlee2008}.

After solving the system \eqref{system} by forward substitution, the coefficients $\alpha_k$ have to be inserted in the representation formula \eqref{solBarrier}\footnote{$\textrm{ceil}[\cdot]$:=function that rounds its argument to the nearest integers towards plus infinity.}
\begin{equation}\label{postpro}
\begin{array}{l}
\displaystyle u(x,\tau)\approx \int_{-\infty}^{B}u_0(y)G(y,0;x,\tau)dy+\sum_{k=1}^{\textrm{ceil}[\frac{\tau}{\Delta t}]}\alpha_k\int_{t_{k-1}}^{\min(t_{k},\tau)}\frac{{\sigma}^2}{2}G(B,s;x,\tau)ds=\vspace{0.2cm}\\
\displaystyle=
\int_{a}^{\min(B,\log(E),b)}(E-e^y)\left\{\sum_{n=0}^{N_F-1}{}^{\prime}F_n[x,0,\tau]\cos\left(n\pi\frac{y-a}{b-a}\right)\right\}dy+\vspace{0.2cm}\\
\displaystyle +\sum_{k=1}^{\textrm{ceil}[\frac{\tau}{\Delta t}]}\alpha_k\int_{t_{k-1}}^{\min(t_{k},\tau)}
H[B-a]H[b-B]\frac{{\sigma}^2}{2}
\left\{\sum_{n=0}^{N_F-1}{}^{\prime}F_n[x,s,\tau]\cos\left(n\pi\frac{B-a}{b-a}\right)\right\}ds\,.

\end{array}
\end{equation}
with $F_n$ defined as in \eqref{F_k} now fixing
\begin{equation}\label{bounds_bis}
\begin{array}{l}
\displaystyle a=\int_s^{\tau}\overline{r}(v)dv-\frac{\sigma^2}{2}\left(\tau-s\right)-L\sqrt{\sigma^2\left(\tau-s\right)}+x\vspace{0.2cm}\\
\displaystyle b=\int_s^{\tau}\overline{r}(v)dv-\frac{\sigma^2}{2}\left(\tau-s\right)+L\sqrt{\sigma^2\left(\tau-s\right)}+x
\end{array}\,.
\end{equation}
The option price is finally recovered by the relation \eqref{Change}: $\forall S\in(0,S_B)\,,\,\,\forall t\in [0,T)$
\begin{equation}\label{transformback}
V(S,t)=u(\log(S),T-t)e^{-\int_t^T r(t')dt'}\,.
\end{equation}

\noindent \textbf{Remark.} The application of the COS method in combination with BEM to the BS model is not computationally advantageous in terms of CPU time, as in this case the fundamental solution is analytically known; however, we think that it provides a good simple example to make the procedure clear to the reader and to highlight the details of the implementation.

\subsection{Series truncation error: choice of $N_F$ in Black-Scholes model}\label{sec:COSerroranalysis}
In what follows, we give an example on the estimation of $N_F$ when pricing a plain vanilla call option with current stock price $S_0=100$, strike $E=120$, time to maturity $T=0.1$, risk-free interest rate $r=0.05$ and volatility $\sigma=0.2$, having set $[a,b]$ as in \eqref{bounds_bis} with $L=10$. We select the BS model for the underlying, since the exact price of the option, which serves as the reference value, is given by the celebrated BS formulae. As pointed out in \cite{FangOosterlee2008}, the pricing of the option under the BS model, does not introduce any other numerical error rather than those derived from the Fourier inversion, which is the only step that we use in our barrier option pricing problem. The selection of a short maturity for the numerical experiment is motivated by the fact that the Fourier inversion step in our barrier option pricing problem involves short time intervals.

When the asset dynamics is governed by the BS model, then $\ln \left(\frac{S_T}{E} \right)$ is normally distributed with mean $\ln \left(\frac{S_0}{E}+\left(r-\frac{1}{2}\sigma^2 \right)T \right)$ and variance $\sigma^2T$. The modulus of the characteristic function of the log-asset price $\ln \left(\frac{S_T}{E} \right)$ reads,
$$|\mathcal{G}(\omega)|=e^{-\frac{1}{2}\sigma^2T \omega^2},$$ 
and, from \eqref{mod},
\begin{equation}\label{eq:seriesBS}
\bar{\epsilon}_t \le \frac{2}{b-a}\sum_{n=N_F}^{\infty} e^{-\frac{1}{2}\left(\frac{\sigma\pi}{b-a}\right)^2Tn^2}.
\end{equation}
The series in expression (\ref{eq:seriesBS}) is convergent, since,
\begin{equation}\label{eq:seriesconverge}
\bar{\epsilon}_t \le \frac{2}{b-a}\sum_{n=N_F}^{\infty} e^{-\frac{1}{2}\left(\frac{\sigma\pi}{b-a}\right)^2Tn^2} \le \frac{2}{b-a}\sum_{n=N_F}^{\infty} e^{-dn}=\frac{2}{b-a}\cdot \frac{e^{-dN_F}}{1-e^{-d}},
\end{equation}
where $d=\frac{1}{2}\left(\frac{\sigma\pi}{b-a}\right)^2T$. 
The error bound of expression (\ref{eq:seriesconverge}) is not sharp enough to obtain an appropriate estimation of $N_F$ and it is solely given to prove the convergence of the infinite series in expression (\ref{eq:seriesBS}). Since the terms of the series in expression (\ref{eq:seriesBS}) decrease very rapidly, we determine $N_F$ by means of the first term of the series. Given a tolerance error $\epsilon_{N_F}$, we select the smallest value  $N_F$ satisfying,
\begin{equation}
    \frac{2}{b-a} e^{-\frac{1}{2}\left(\frac{\sigma\pi}{b-a}\right)^2TN_F^2} \le \epsilon_{N_F},
\end{equation}
that is,
\begin{equation}\label{eq:estimateNF}
    N_F=\left \lceil \sqrt{\frac{1}{T}\left(\frac{b-a}{\sigma \pi} \right)^2 \ln \left(\frac{b-a}{2}\epsilon_{N_F} \right)}\right \rceil,
\end{equation}
where $\lceil x \rceil:=\min \{ k \in \mathbb{Z}: k \ge x \}$. For instance, when $\epsilon_{N_F}=10^{-3}$ then expression (\ref{eq:estimateNF}) gives $N_F=25$.

To support the validity of the choice of $N_F$, we compute the prices of the call option by means of the COS method when $N_F=20,25,30,35,40,45,50$, and we compare them with the reference price. The absolute errors corresponding to each value of $N_F$ are plotted in red color in Figure \ref{Fig:N_terms_COS}. The blue points in Figure \ref{Fig:N_terms_COS} represent the error bound given in expression (\ref{eq:seriesBS}). 
Finally, the black points represent only the first term in the series (\ref{eq:seriesBS}) multiplied by $2/(b-a)$. We observe that the error increases slightly when adding many terms in the series with respect to the consideration of only the first term, but the order of magnitude remains the same. Further, we show that the a priori estimation of $N_F$ is accurate, in accordance with the a posteriori absolute errors obtained with respect to the reference value.

\begin{center}
	\includegraphics[trim=500 5 500 0, clip,scale=0.5]{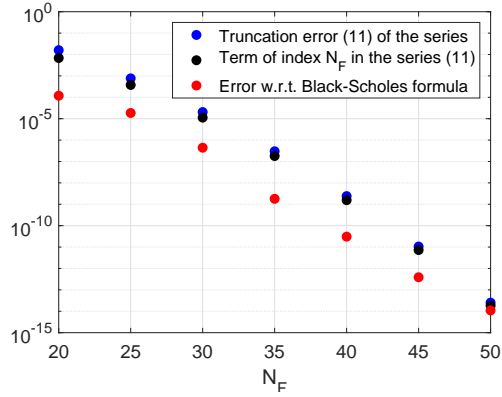}
	\captionof{figure}{Series truncation error. The absolute errors are calculated with respect to the reference value given by the BS formula.}\label{Fig:N_terms_COS}
\end{center}
A similar argument can be carried out for any other process for the log-asset price, provided that we know its characteristic function.

\subsection{Numerical results in the BS framework}
\label{sec;NumExBS}
In this section, we briefly want to show the performance of the procedure coupling the COS method with the BEM for the evaluation of barrier options in the BS framework, keeping in mind that this is only an introductory example to illustrate the methodology in a simple mathematical setting. Our reference for comparison is the implementation of the BEM without the use of the COS method, as developed in \cite{Guardasoni2018}.

First of all, we highlight that the two numerical approaches produce exactly the same option prices, providing accurate approximations for very low levels of the time discretization parameter $N_{\Delta t}$. To give evidence of that, Table \ref{Tab:r_cputime} shows the approximate option price $V(35,0)$ provided by both methods as $N_{\Delta t}=2^{n_{\Delta t}}$ increases, using the parameter values listed in Table \ref{Tab:r_data}.
Moreover, although slightly more efficient, the COS BEM procedure has substantially the same computational cost of the one in \cite{Guardasoni2018}.
$$
\begin{array}{|c|c|c|c|c|c|c|c|}
  \hline
  S_u & t_0 & T & S & E & r(t) & d & \sigma \\
  \hline
  40 & 0 & 1 & 35 & 50 &\left\{\begin{array}{ll}
                                  r_1=0.01 & t<0.25 \\
                                  r_2=0.03 & 0.25\leq t\leq T
                                \end{array}\right. & 0
                                & 0.105 \\
  \hline
\end{array}\vspace{-0.2cm}
$$\captionof{table}{BS model. Up-and-out put option data.}\vspace{-0.0cm}\label{Tab:r_data}

$$
\begin{array}{l|c|c|c}
 n_{\Delta t} & V(35,0) & \textrm{CPU time-BEM \cite{Guardasoni2018}} & \textrm{CPU time-COS BEM}\\
 \hline
4  & 11.43996 & 1.1E+00 & 1.2E+00 \\
8  & 11.43862 & 2.0E+00 & 3.6E+00 \\
16 & 11.43811 & 5.7E+00 & 3.8E+00 \\
32 & 11.43789 & 1.4E+01 & 1.1E+01 \\
64 & 11.43781 & 3.7E+01 & 2.7E+01 \\
\end{array}$$
\captionof{table}{Value and CPU time (in seconds) for the put up-and-out option price at $S=35$ and $t=0$ obtained by the BEM Matlab code (on the left) of \cite{Guardasoni2018} and by the new COS BEM Matlab code (on the right) with $\Delta t=T/2^{n_{\Delta t}}$.}\label{Tab:r_cputime}\vspace{0.0cm}

\section{Heston model}\label{sec:Heston}
\subsection{Down-and-Out Call Option}

In the Heston model the option price $V$ depends on three independent variables  $V(x,v,t)$: the log-asset value $x=\log S$, the variance or squared volatility $v$ and the time $t$. 

Let us consider the undiscounted price $u(x,v,t)=V(x,v,t) e^{r(T-t)}$ of a down-and-out call option, i.e an option that vanishes if the underlying asset decreases enough to reach an assigned lower barrier $S_B=e^B$ before the expiry date $T$. 

In this framework, the involved differential model problem is
\begin{eqnarray}
                  \label{HestonPDE}\displaystyle\frac{\partial u}{\partial t}+\frac{1}{2}v\frac{\partial^2 u}{\partial x^2}+\rho \eta v \frac{\partial^2 u}{\partial x \partial v}+\frac{1}{2}\eta^2 v
\frac{\partial^2 u}{\partial v^2} + \left(r-\delta-\frac{1}{2}v\right) \frac{\partial u}{\partial x}- (\lambda(v-\bar v)-\theta v) \frac{\partial u}{\partial v}=0, \hspace{1.cm}\\
                  \displaystyle x\in\Omega_x:=(B,+\infty)\,,\, v\in \Omega_v:=(0,+\infty)\,,\,t\in[0,T]\,\quad\nonumber\\\nonumber\\
u(x,v,T)=(e^{x}-E)^+=\max(e^x -E,0)\hspace{5.cm}x\in\Omega_x\,,\, v\in \Omega_v\quad\\
u(B,v,t)=0\hspace{8.9cm} v\in\Omega_v\,,\,t\in[0,T]\quad\label{BarrierCond_H}
\end{eqnarray}
The parameter $\lambda$ is the speed of mean reversion, $\bar v$ is the mean level of variance, $\eta$ is the volatility of volatility, $\delta$ is the dividend yield and $\theta$, the market price of volatility risk, is assumed to be $0$. We assume that the Feller condition $2\lambda \bar v\geq \eta^2$ holds.\\

$\bullet$ In \cite{GuardasoniSanfelici2016}, it is proved that $u$ can be described by the following integral representation formula
\begin{equation}\label{BatesForm1}
\begin{array}{r}
\displaystyle u(x,v,t)=\int_{\Omega_x}\int_{\Omega_v} (e^y-E)^+ G(y,w,T;x,v,t)dw\,dy-\!\int_t^T\!\! \int_{\Omega_v}  \frac{w}{2}G(B,w,s;x,v,t) \frac{\partial u}{\partial y}(B,w,s) dw\, ds\,\vspace{0.2cm}\\\displaystyle
x\in\Omega_x\,,\,v\in \Omega_v\,,\,t\in[0,T]
\end{array}
\end{equation}
where $G$ is the fundamental solution of \eqref{HestonPDE}, i.e. the joint transition probability density function of the underlying stochastic process, for which an explicit expression is not available. In the same article, an implicit expression is given 
\begin{equation}G(y,w,s;x,v,t)=p_v(w, s-t|v) p(y-x, s-t | w , v).\label{DensBates}\end{equation}

The transition density of the variance $w$ conditioned on $v$ is given by
\begin{equation}\label{pv}
	p_v(w, s-t|v)=ce^{-b-q}\left(\frac{q}{b}\right)^{\frac{a-1}{2}}I_{a-1}(2\sqrt{bq}),
\end{equation}
where $c=2\lambda/((1-e^{-\lambda (s-t)})\eta^2)$, $b=cve^{-\lambda(s-t)}$, $q=cw$, $a=2\lambda \bar v/ \eta^2$ and $I_a(q)$ is the modified Bessel function of the first kind.

Moreover, the transition density function $p(z, s-t | w , v)$ of the logarithm of the stock price given $v$ and given $w$ is known through its Fourier transform w.r.t. the variable $z=y-x$ 
\begin{equation}\label{CharBates}
\begin{array}{rcl}
\displaystyle\hspace{-0.5cm}\hat p(\omega;v,w,s-t)=\int_{-\infty}^{\infty}p(z,s-t|w,v)e^{i\omega z}dz \!\!\!&\!\!=\!\!&\!\!\displaystyle\exp \left\{ \mathbf{i}\omega (r-\delta)(s-t)+\frac{\rho}{\eta}(w-v-\lambda \bar v(s-t)) \right\} \vspace{0.2cm}\\
		\!\!&\!\!\times\!\!&\!\!\displaystyle\Phi\left( \omega \left( \frac{\lambda \rho}{\eta}-\frac{1}{2} \right) +\frac{\omega^2}{2} \mathbf{i} (1-\rho^2) \right),\end{array}\end{equation}
where $\Phi(a)$ is the characteristic function of the integrated variance $\int_{t}^s v(\tau) d\tau$ given $v$ and $w$:\\
\begin{equation}\label{CharFunVar}\begin{array}{l}\displaystyle\Phi(a)=\frac{\gamma(a)e^{-\frac{1}{2}(\gamma(a)-\lambda)(T-t_0)}(1-e^{-\lambda(T-t_0)})}
		{\lambda(1-e^{-\gamma(a)(T-t_0)})}\vspace{0.2cm}\\
		\displaystyle\times \exp \left\{ \frac{v_0+v_T}{\eta^2}\left[ \frac{\lambda (1+e^{-\lambda(T-t_0)})}{(1-e^{-\lambda(T-t_0)})}-\frac{\gamma(a)(1+e^{-\gamma(a)(T-t_0)})}{(1-e^{-\gamma(a)(T-t_0)})} \right] \right\}\vspace{0.2cm}\\
		\displaystyle\times \frac{I_{\frac{1}{2}d-1} \left( \sqrt{v_0v_T} \frac{4\gamma(a)e^{-\frac{1}{2}\gamma(a)(T-t_0)}}{\eta^2(1-e^{-\gamma(a)(T-t_0)})} \right)}{I_{\frac{1}{2}d-1}  \left( \sqrt{v_0v_T} \frac{4\lambda e^{-\frac{1}{2}\lambda(T-t_0)}}{\eta^2(1-e^{-\lambda(T-t_0)})} \right)},\end{array}\end{equation}
with $\gamma(a)=\sqrt{\lambda^2-2\eta^2\mathbf{i}a}$ and $d=4\bar v \lambda/\eta^2$.
Therefore, the transition density function $p(z, s-t | w , v)$ will be recovered from its characteristic function by the COS method.

Finally, in our computations, we will use the characteristic function also for the log-price only
\begin{equation*}\begin{array}{c}
		\displaystyle \phi_s(\omega;x,v,t)=\exp \left\{ \mathbf{i}\omega (r-\delta)(s-t)+\frac{v}{\eta^2}\left( \frac{1-e^{-D(s-t)}}{1-Ce^{-D(s-t)}} \right)(\lambda-\rho \eta \mathbf{i} \omega -D)+ \right. \vspace{0.2cm}\\
		\displaystyle\left. +\frac{\lambda\bar v}{\eta^2}\left( (s-t)(\lambda-\rho \eta \mathbf{i} \omega -D)-2\log \left( \frac{1-Ce^{-D(s-t)}}{1-C} \right) \right) +\mathbf{i}\omega x \right\},
	\end{array}
\end{equation*}
where $D=\sqrt{(\lambda-\rho \eta \mathbf{i} \omega)^2+(\omega^2+\mathbf{i}\omega)\eta^2}$ and $C=\frac{\lambda-\rho \eta \mathbf{i} \omega -D}{\lambda-\rho \eta \mathbf{i} \omega +D}$. From its Fourier inversion, it is possible to get the marginal PDF $\widetilde{G}$ that has the following relation with $G$
\begin{equation}\label{Gtilde}
\begin{array}{l}
    \hspace{-0.45cm}\displaystyle \widetilde{G}(y,s;x,v,t):=\int_{\Omega_v}\!\!  G(y,w,s;x,v,t) dw ={\cal F}_\omega^{-1}[\phi_s](y,s;x,v,t)=\frac{1}{2\pi}\int_{-\infty}^{+\infty}\phi_s(\omega;x,v,t)e^{-\mathbf{i}\omega y}d\omega.
\end{array}
\end{equation}
Also (\ref{Gtilde}) can be approximated by means of the COS method.\\

$\bullet$ Letting $x\rightarrow B$ in \eqref{BatesForm1} and taking into account the vanishing condition \eqref{BarrierCond_H} for the option price at the log-barrier $B$, we obtain the BIE
\begin{equation}\label{BIE_H}
\begin{array}{l}
\displaystyle 0=u(B,v,t)= \!\! \int_{\Omega_x}\int_{\Omega_v}\!\!\max(e^y-E,0)G(y,w,T;B,v,t)dw\,dy+\vspace{0.2cm}\\
\displaystyle -\int_{t}^T\int_{\Omega_v}\frac{w}{2}G(B,w,\tau;B,v,t)\frac{\partial u}{\partial y}(B,w,\tau)dw\,d\tau
\end{array}
\end{equation}
whose sole unknown is the function $\frac{\partial u}{\partial y}(B,w,\tau)$.

Then, \eqref{BIE_H} is numerically solved, obtaining an approximation of $\frac{\partial u}{\partial y}(B,w,\tau)$ that, inserted in \eqref{BatesForm1}, provides an approximation for the solution $u$ of the differential problem \eqref{HestonPDE}-\eqref{BarrierCond_H}, wherever in the domain $\Omega_x\times\Omega_v$, at any instant in $[0,T]$.

\subsection{Approximation of the Heston BIE solution by the COS method}
\label{sec;NumApproxBIE_H}
The time interval $[0,T]$ is subdivided in $N_{\Delta t}\in\mathbb{N}^+$ intervals of length $\Delta t=T/N_{\Delta t}$, 
$$
\displaystyle t_{k}=k\Delta t,\quad k=0,\ldots,N_{\Delta t}\,
$$
and, in time, the unknown $\frac{\partial u}{\partial y}(B,w,\tau)$ is approximated by piecewise constant basis functions $\varphi_k(\tau),\,k=1,\ldots,N_{\Delta t}$.

With respect to volatility, we know that as it approaches infinity, the price approaches a steady state. So, we can infer that there exists a value $v_\textrm{MAX}$ such that integrals with kernel $G$ involved in our discretization algorithm over interval $[v_\textrm{MAX},+\infty]$ are negligible (in our numerical examples we set $v_\textrm{MAX}=2*\max(v,\overline{v})$); then we can introduce a uniform decomposition of the truncated variance domain $[0,v_\textrm{MAX}]$ in $N_{\Delta v}\in\mathbb{N}^+$ variance intervals of length $\Delta v:=v_\textrm{MAX}/N_{\Delta v}$ 
$$
\displaystyle v_{h}=h\Delta v,\quad v=0,\ldots,N_{\Delta v}\,
$$
and we can choose piecewise constant shape functions $\psi_h(w),\, h=1,\ldots,N_{\Delta v}$ for the approximation in variance of the unknown function
\begin{equation}\label{approx_H}
\displaystyle \frac{\partial u}{\partial y}(B,w,\tau)\approx q(w,\tau):=\sum_{h=1}^{N_{\Delta v}}\sum_{k=1}^{N_{\Delta t}}\alpha_{h}^{(k)}\psi_h(w)\varphi_k(\tau)\,.
\end{equation}
Alternatively, we could resort to an infinite element approach \cite{Sanfelici2004} thus avoiding any domain truncation.\\
Then, after the substitution of $q(w,\tau)$, equation \eqref{BIE_H} is evaluated at the collocation points $(\overline{v}_i,\overline{t}_j)$ choosing, as collocation points, the barycentres of the intervals $[v_{i-1},v_i]\times[t_{j-1},t_j]$
$$\overline{v}_i=\frac{v_i+v_{i-1}}{2},\,i=1,\ldots,N_{\Delta v}\qquad\overline{t}_j=\frac{t_j+t_{j-1}}{2},\,j=1,\ldots,N_{\Delta t}$$
and so obtaining a linear system 
\begin{equation}\label{system_H}
\mathcal{A}\alpha=\mathcal{F}\,,
\end{equation}
whose unknowns are the coefficients of (\ref{approx_H}) collected in the vector $\alpha=\big(\alpha^{(k)}\big|_{k=1,\ldots,N_{\Delta t}}\big)=\big((\alpha_{h}^{(k)}|_{h=1,\ldots,N_{\Delta v}})\big|_{k=1,\ldots,N_{\Delta t}}\big)$
and where
\begin{equation}\label{matrix_H}\displaystyle\mathcal{A}_{ih}^{(jk)}=\int_{\overline{t}_j}^T\int_{\Omega_v}
\frac{w}{2}G(B,w,\tau;B,\overline{v}_i,\overline{t}_j)\psi_h(w)\varphi_k(\tau)dw\,d\tau
\end{equation}
and
\begin{equation}\label{rhs_H}
\begin{array}{rcl}
\mathcal{F}_i^{(j)}&=&\displaystyle\int_{\Omega_x}\int_{\Omega_v}\max(e^y-E,0)G(y,w,T;B,\overline{v}_i,\overline{t}_j)dw\,dy=\vspace{0.2cm}\\
 &=&\displaystyle\int_{\Omega_x}\max(e^y-E,0)\widetilde{G}(y,T;B,\overline{v}_i,\overline{t}_j)dy,
\end{array}
\end{equation}
with $\widetilde{G}$ defined as in \eqref{Gtilde}.\\
The matrix $\mathcal{A}$ defined by \eqref{matrix_H} has a block upper triangular Toeplitz structure in time with $N_{\Delta t}$ blocks, each of dimension $N_{\Delta v}\times N_{\Delta v}$
\begin{equation}\label{Toeplitz_H}
\left(\begin{array}{ccccc}
                \mathcal{A}^{(0)} & \mathcal{A}^{(1)} & \mathcal{A}^{(2)} & \cdots & \mathcal{A}^{(N_{\Delta t}-1)} \\
                 0 & \mathcal{A}^{(0)} & \mathcal{A}^{(1)} & \cdots & \mathcal{A}^{(N_{\Delta t}-2)} \\
                 0 & 0 & \mathcal{A}^{(0)} & \ddots & \vdots \\
                \vdots & \vdots & \ddots & \ddots & \mathcal{A}^{(1)} \\
                0 & 0 & \cdots & 0 & \mathcal{A}^{(0)} \\
              \end{array}
            \right)
\left(\begin{array}{c}
                \alpha^{(1)} \\
                \alpha^{(2)} \\
                \alpha^{(3)} \\
                \vdots \\
                \alpha^{(N_{\Delta t})} \\
              \end{array}
            \right)=\left(\begin{array}{c}
                \mathcal{F}^{(1)} \\
                \mathcal{F}^{(2)} \\
                \mathcal{F}^{(3)} \\
                \vdots \\
                \mathcal{F}^{(N_{\Delta t})} \\
              \end{array}
            \right)
\end{equation}
since its elements depend on the difference $k-j=\ell,\, \ell=0,\ldots,N_{\Delta t}-1$ and reduce to
\begin{equation}
\begin{array}{l}
\displaystyle\mathcal{A}_{ih}^{(jk)}=\!\!\int_{\max(\overline{t}_j,t_{k-1})}^{t_k}\hspace{-0.7cm}H[t_k-\max(\overline{t}_j,t_{k-1})]\int_{v_{h-1}}^{v_{h}}
\frac{w}{2}G(B,w,\tau;B,\overline{v}_i,\overline{t}_j)dw\,d\tau=\vspace{0.2cm}\\
\displaystyle=\!\!\int_{\max(\overline{t}_j,t_{k-1})}^{t_k}\hspace{-0.7cm}H[t_k-\max(\overline{t}_j,t_{k-1})]\int_{v_{h-1}}^{v_{h}}
\frac{w}{2}p_{v}(w,\tau-\overline{t}_j|\overline{v}_i)p(0,\tau-\overline{t}_j|w,\overline{v}_i)dw\,d\tau=\vspace{0.2cm}\\
\displaystyle=\!\!\int_{\frac{1}{2}-\frac{1}{2}H[\ell]}^{1}\int_{v_{h-1}}^{v_{h}}
\!\!\frac{\Delta t}{2}w\,p_{v}(w,\Delta t(\ell-\frac{1}{2}+s)|\overline{v}_i)p(0,\Delta t(\ell-\frac{1}{2}+s)|w,\overline{v}_i)dw ds=:\mathcal{A}^{(\ell)}_{ih}
\end{array}
\end{equation}
as, with the change of variable $\tau=\Delta t(k+s-1)$, we get $\tau-\overline{t}_j=\Delta t(k-j-\frac{1}{2}+s)=\Delta t(\ell-\frac{1}{2}+s)$.\\
To recover the function $p$ in the integrand, we again apply the COS method for the fast inverse transform of the function \eqref{CharBates} and get 
\begin{equation}\label{matrix_el_H}
\begin{array}{rcl}
\displaystyle\mathcal{A}^{(\ell)}_{ih}&=&\displaystyle\!\!\int_{\frac{1}{2}-\frac{1}{2}H[\ell]}^{1}\int_{v_{h-1}}^{v_{h}}
\!\!\frac{\Delta t}{2}w\,p_{v}(w,\Delta t(\ell-\frac{1}{2}+s)|\overline{v}_i)\!\!\vspace{0.2cm}\\
&&\displaystyle\left\{\sum_{n=0}^{N_F-1}{}^{\prime}F_n[\overline{v}_i,w,\Delta t(\ell-\frac{1}{2}+s)]\cos\left(n\pi\frac{-a}{b-a}\right)\right\} dw ds
\end{array}
\end{equation}
with
\begin{equation}\label{F_k_H_matrix}
\begin{array}{l}
\displaystyle F_n[v,w,t]=\frac{2}{b-a}\textrm{Re}\left\{\widehat{p}\left(\frac{n\pi}{b-a};v,w,t\right)e^{-\mathbf{i}n\pi\frac{a}{b-a}}\right\},\qquad n=0,\ldots,N_F-1,
\end{array}
\end{equation}
where, for the bounds, we consider the same ones suggested in \cite{FangOosterlee2008}:
\begin{equation}\label{bounds_H_matrix}
\begin{array}{l}
\displaystyle a=c_1-L\sqrt{c_2}\qquad\displaystyle b=c_1+L\sqrt{c_2}\vspace{0.2cm}\\
\displaystyle c_1=(r-\delta)t+\left(1-e^{-\lambda t}\right)\frac{\overline{v}-\overline{v}_i}{2\lambda}-\frac{1}{2}\overline{v}t\vspace{0.2cm}\\
\displaystyle c_2=\frac{1}{8\lambda^3}\left\{\eta t\lambda e^{-\lambda t}(\overline{v}_i-\overline{v})(8\lambda\rho-4\eta)\right.\vspace{0.2cm}\\
\hspace{1.7cm}+\lambda\rho\eta\left(1-e^{-\lambda t}\right)(16\bar{v}-8\bar{v}_i)\vspace{0.2cm}\\
\hspace{1.7cm}+2\bar{v}\lambda t(-4\lambda\rho\eta+\eta^2+4\lambda^2)\vspace{0.2cm}\\
\hspace{1.7cm}+\eta^2\big((\bar{v}-2\bar{v}_i)e^{-2\lambda t}+\bar{v}(6e^{-\lambda t}-7)+2\bar{v}_i\big)\vspace{0.2cm}\\
\hspace{1.7cm}\left.+8\lambda^2(\bar{v}_i-\bar{v})(1-e^{-\lambda t}\right\}
\end{array}
\end{equation}
Based on the error analysis carried out int the following Section \ref{sec:COSerroranalysis_H}, the parameters of the cosine expansion in all the simulations are set as $L=30, N_F=128$.\\

\textbf{Remark.} In \cite{FangOosterlee2008}, the authors apply the COS method to the marginal PDF $\tilde{G}$ in \eqref{Gtilde}, but here this strategy is applicable only to the right-hand side entries, as shown below.\\

\textbf{Remark.} With the application of the COS method we are able to finally overcome in an efficient way the bottleneck of the Fourier inversion that was already faced in our previous paper \cite{GuardasoniSanfelici2016}. In that paper we used the Gauss-Kronrod adaptive quadrature implemented in the Matlab function \verb+quadgk+. The well known strategy of P. Carr and D.B. Madan illustrated in \cite{Carr1999} is not suitable for our algorithm because it provides the inverse transform in a set of points and not only at one specific point as necessary in our integrals; moreover, numerically, it has proved to be very sensitive to the variation of the tuning parameters involved and it does not allow us to achieve the required accuracy.\\ 

For the computation of the right-hand side term, we can exploit the relations in \eqref{Gtilde} and use the COS method to get
\begin{equation}\label{rhs_el_H}
\begin{array}{rcl}
\mathcal{F}_i^{(j)} \!\!&=&\!\! \displaystyle\int_{B}^{+\infty}\max(e^y-E,0)\widetilde{G}(y,T;B,\overline{v}_i,\overline{t}_j)dy\vspace{0.2cm}\\
\!\!&=&\!\! \displaystyle\int_{\max\big(B,\log(E)\big)}^{+\infty}(e^y-E)\widetilde{G}(y,T;B,\overline{v}_i,\overline{t}_j)dy\vspace{0.2cm}\\
\!\!&=&\!\! \displaystyle H[b-\max\big(B,\log(E)\big)]\int_{\max\big(a,B,\log(E)\big)}^{b}\!\!\!\!\!\!\!(e^y-E)\sum_{n=0}^{N_F-1}{}^{\prime}F_n[B,\overline{v}_i,T,\overline{t}_j]\cos\left(n\pi\frac{y-a}{b-a}\right) dy\vspace{0.2cm}\\
\!\!&=&\!\! \displaystyle H[b-\max\big(B,\log(E)\big)]
\sum_{n=0}^{N_F-1}{}^{\prime}F_n[B,\overline{v}_i,T,\overline{t}_j]V_n
\end{array}
\end{equation}
with
\begin{equation}\label{F_k_H_rhs}
\begin{array}{l}
\displaystyle F_n[x,v,T,t]=\frac{2}{b-a}\textrm{Re}\left\{\phi_T\left(\frac{n\pi}{b-a};x,v,t\right)e^{-\mathbf{i}n\pi\frac{a}{b-a}}\right\},\qquad n=0,\ldots,N_F-1,\vspace{0.2cm}\\
\displaystyle V_n=\int_{\max\big(a,B,\log(E)\big)}^{b}\!\!\!\!\!(e^y-E)\cos\left(n\pi\frac{y-a}{b-a}\right)dy, \qquad n=0,\ldots,N_F-1,\quad \textrm{analitically integrated}
\end{array}
\end{equation}
and bounds:
\begin{equation}\label{bounds_H_rhs}
\begin{array}{l}
\displaystyle a=c_1-L\sqrt{c_2}+x\qquad\displaystyle b=c_1+L\sqrt{c_2}+x\vspace{0.2cm}\\
\displaystyle c_1=(r-\delta)(T-t)+\left(1-e^{-\lambda(T-t)}\right)\frac{\overline{v}-\overline{v}_i}{2\lambda}-
\frac{1}{2}\overline{v}(T-t)\vspace{0.2cm}\\
\displaystyle c_2=\frac{1}{8\lambda^3}\left\{\eta(T-t)\lambda e^{-\lambda(T-t)}(\overline{v}_i-\overline{v})(8\lambda\rho-4\eta)\right.\vspace{0.2cm}\\
\hspace{1.7cm}+\lambda\rho\eta\left(1-e^{-\lambda(T-t)}\right)(16\bar{v}-8\bar{v}_i)\vspace{0.2cm}\\
\hspace{1.7cm}+2\bar{v}\lambda(T-t)(-4\lambda\rho\eta+\eta^2+4\lambda^2)\vspace{0.2cm}\\
\hspace{1.7cm}+\eta^2\big((\bar{v}-2\bar{v}_i)e^{-2\lambda(T-t)}+\bar{v}(6e^{-\lambda(T-t)}-7)+2\bar{v}_i\big)\vspace{0.2cm}\\
\hspace{1.7cm}\left.+8\lambda^2(\bar{v}_i-\bar{v})(1-e^{-\lambda(T-t)})\right\}
\end{array}
\end{equation}

Once all the elements of the linear system (\ref{Toeplitz_H}) have been evaluated, due to the particular structure of the matrix $\mathcal{A}$, the approximate solution $q(w,\tau)$ of the BIE (\ref{BIE_H}), expressed by the vector of coefficients $\alpha$ in (\ref{approx_H}), can be obtained by block-backward substitution: the only matrix to be inverted is the diagonal block $\mathcal{A}^{(0)}$, the others update at every time step the right-hand side.\\
At the end, the undiscounted price of barrier option $u(x,v,t)$ in $\Omega_x\times\Omega_v$ for $t\in[0,T)$ is obtained by introducing $q(w,\tau)$ into equation (\ref{BatesForm1})\footnote{$\textrm{floor}[\cdot]$:=function that rounds its argument to the nearest integers towards minus infinity.}:
\begin{equation}\label{postpro_H}
\begin{array}{l}
\displaystyle u(x,v,t)\approx \int_{B}^{+\infty}\max(e^y-E,0)\widetilde{G}(y,T;x,v,t)dy+\vspace{0.2cm}\\
\displaystyle -\sum_{h=1}^{N_{\Delta v}}\sum_{k=\textrm{floor}[\frac{t}{\Delta t}]+1}^{N_{\Delta t}}\alpha_h^{(k)}\int_{\max(t,t_{k-1})}^{t_k}\int_{v_{h-1}}^{v_h}\frac{w}{2}G(B,w,\tau;x,v,t)dw\,d\tau\,.
\end{array}
\end{equation}
The manipulation of the first term on the right-hand side of (\ref{postpro_H}) is developed as for the elements ${\cal F}_i^{(j)}$ of equations \eqref{rhs_el_H}-\eqref{F_k_H_rhs}):
\begin{equation}\label{postpro_initdata_H}
\begin{array}{l}
    \displaystyle \int_{B}^{+\infty}\max(e^y-E,0)\widetilde{G}(y,T;x,v,t)dy=H[b-\max\big(B,\log(E)\big)]
\sum_{n=1}^{N_F-1}{}^{\textcolor[rgb]{1,0,0}{\prime}} F_n[x,v,T,t]V_n
\end{array}
\end{equation}
with bounds $[a,b]$ as in \eqref{bounds_H_rhs}.

If the time of the option price evaluation is $t=0$, the second term on the right-hand side of (\ref{postpro_H}) for the evaluation of the option price $u(x,v,0)$ is
\begin{equation}\label{postpro_intsol_H}
\begin{array}{l}
    \displaystyle -\sum_{h=1}^{N_{\Delta v}}\sum_{k=1}^{N_{\Delta t}} \alpha_h^{(k)}\int_{t_{k-1}}^{t_k}\int_{v_{h-1}}^{v_h}\frac{w}{2}G(B,w,\tau;x,v,0)dw\,d\tau=\vspace{0.2cm}\\
\displaystyle=-\sum_{h=1}^{N_{\Delta v}}\sum_{k=1}^{N_{\Delta t}} \alpha_h^{(k)}\int_{t_{k-1}}^{t_k}\int_{v_{h-1}}^{v_{h}}
\frac{w}{2}p_v(w,\tau|v)p(B-x,\tau|w,v)dw\,d\tau=\vspace{0.2cm}\\
\displaystyle=-\sum_{h=1}^{N_{\Delta v}}\sum_{k=1}^{N_{\Delta t}} \alpha_h^{(k)}\int_{0}^{1}\int_{v_{h-1}}^{v_{h}}
\!\!\frac{\Delta t}{2}w\,p_{v}(w,\Delta t(k-1+s)|v)\vspace{0.2cm}\\
\displaystyle\quad\sum_{n=0}^{N_F-1}{}^{\prime}F_n[v,w,\Delta t(k-1+s)]\cos\left(n\pi\frac{B-x-a}{b-a}\right) dw\, ds
\end{array}
\end{equation}
and it can be computed as done for the linear system entries by \eqref{F_k_H_matrix} and \eqref{bounds_H_matrix}. Here, we observe that, in the evaluation of the cosine expansion, for $t\rightarrow 0$, $c_1$ and $c_2$ in \eqref{bounds_H_matrix} tend to vanish and, therefore, a value of  $L$ larger than in the evaluation of matrix entries is required to avoid collapse of the interval $[a,b]$ around point $z$ ($L=30$ is sufficient for all integrals).

Finally to get the price of the contingent claim we apply the relation $V(S,v,0)=u(x,v,0)e^{-rT}$.

\subsection{Series truncation error: choice of $N_F$ in Heston model}\label{sec:COSerroranalysis_H}
Note that in the evaluation of barrier options the fundamental solution is given not through the characteristic function but by formula \eqref{BatesForm1} where the Fourier transform of the transition density function $p(z,s-t|w,v)$ is involved 
$$
p(z,s-t|w,v)=\mathcal{F}_\omega^{-1}[\hat{p}(\omega;v,w,s-t)](z).
$$
Starting from the error bound \eqref{mod},  the a priori choice of $N_F$ must be done on
\begin{equation}\label{mod_H}
\bar{\epsilon}_t \le \frac{2}{b-a}\sum_{n=N_F}^{\infty}\left|\hat{p}\left(\frac{n\pi}{b-a};v,w,s-t\right)\right|.
\end{equation}

To support the validity of the strategy introduced in the BS Section \ref{sec:COSerroranalysis}, we compute, by means of the COS method applied to $\hat{p}$, the prices of the call option without barriers with current stock price $S_0=150$, maturity\footnote{The most challenging task involved in our simulations is the computation of entries in $\mathcal{A}^{(0)}$ defined in \eqref{matrix_el_H}, where we have to accurately integrate the fundamental solution in a time interval of length $\Delta t$. To find $L$ and $N_F$ suitable for these integrals we check them evaluating the accuracy of plain vanilla option price for a short maturity of the order $\Delta t$.} $T=0.05$ and the other parameters as defined in Table \ref{Tab:H_data} choosing $N_F=32:16:128$ and having set $[a,b]$ as in \eqref{bounds_H_matrix} with $L=30$. Then we compare them with the reference price given by the formula in \cite{Heston1993}.
The absolute errors corresponding to each value of $N_F$ are plotted in red color in Figure \ref{Fig:N_terms_COS_H}. The blue points in Figure \ref{Fig:N_terms_COS_H} represent the error bound given in expression (\ref{mod_H}) for $s-t=T$ and $w=0.01$. 
Finally, the black points represent only the first term in the series (\ref{mod_H}) multiplied by $2/(b-a)$. 

\begin{center}
	\includegraphics[trim=500 5 500 0, clip,scale=0.5]{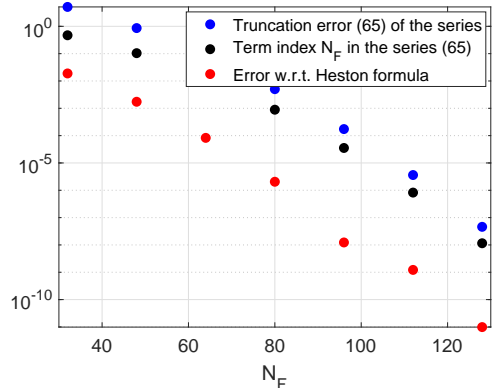}
	\captionof{figure}{Series truncation error. The absolute errors are calculated with respect to the reference value given by the Heston formula.}\label{Fig:N_terms_COS_H}
\end{center}

\subsection{Hedging}
\label{sec;Hedge}
Hedging can be made directly on the representation formula with the post-processing.\\ For example the $\Delta-$greek can be computed without evaluating the option values (as illustrated also in the BS framework in \cite{Guardasoni2018}):

\begin{equation}\label{Delta}
\begin{array}{r}
\displaystyle
\frac{\partial V}{\partial S}(S,v,t)=e^{-r(T-t)}e^{-x}\frac{\partial u}{\partial x}(x,v,t)=e^{-r(T-t)}e^{-x}\left\{\int_{\Omega_x}\int_{\Omega_v} (e^y-E)^+ \frac{\partial G}{\partial x}(y,w,T;x,v,t)dw\,dy\right.\vspace{0.2cm}\\\displaystyle\left.
-\!\int_t^T\!\! \int_{\Omega_v}  \frac{w}{2}\frac{\partial G}{\partial x}(B,w,s;x,v,t) \frac{\partial u}{\partial y}(B,w,s) dw\, ds\right\}\,\vspace{0.2cm}\\\displaystyle
x\in\Omega_x\,,\,v\in \Omega_v\,,\,t\in[0,T]
\end{array}
\end{equation}
and
$$
\begin{array}{r}
\displaystyle
\frac{\partial G}{\partial x}(y,w,s;x,v,t)=p_v(w, s-t|v)\frac{\partial p}{\partial x}(y-x, s-t | w , v)\vspace{0.2cm}\\
\displaystyle=p_v(w, s-t|v)\frac{\partial }{\partial x}\textrm{Re}\left\{\int_{-\infty}^{\infty}\hat p(\omega;v,w,s-t)e^{-i\omega (y-x)}d\omega\right\}\,.
\end{array}
$$

\subsection{Numerical results for a Down-and-Out Call Option}
\label{sec;NumExH}

The efficiency and accuracy of the BEM for multifactor models like Heston’s,
where the analytical expression for the fundamental solution is not available and no closed-form barrier option formulas are known, have already been studied in \cite{GuardasoniSanfelici2016}.
The convergence rate of BEM proved to be very fast in both time and variance.
In fact, the rate of decay of the error benefits from the high order of regularity of the solution
in the interior of the integration domain and exponential convergence was detected with respect to both $N_{\Delta t}$ and $N_{\Delta v}$.
A deep numerical analysis provided evidence of greater accuracy and reduced computational cost in comparison with results achieved by the conditional MC simulation \cite{Glasserman2004} and highlighted BEM as a valid and worthwhile alternative to traditional MC approaches.

We compare here the efficiency in terms of computation time of the BEM code implemented in \cite{GuardasoniSanfelici2016} and the COS BEM code with the addition of the COS method to compute the Fourier inverse transforms. 
The down-and-out call option parameters used in our simulations are listed in Table \eqref{Tab:H_data}.
$$
\begin{array}{|c|c|c|c|c|c|c|c|c|}
  \hline
  \lambda & \bar{v} & \rho & \eta & r & \delta & E & e^B & T \\
  \hline
  4 & 0.04 & -0.5 & 0.1 & 0.05 & 0.02 & 100 & 110 & 1\\
  \hline
\end{array}\vspace{-0.2cm}
$$\captionof{table}{Heston model. Down-and-out call option data.}\label{Tab:H_data}\vspace{0.2cm}

Setting the truncated variance domain equal to $[0,v_{\max}]=[0,0.08]$, the present option prices at $S=115$ and $S=150$ and at the current variance value $v=0.01$ are displayed in Tables \ref{Tab:Heston} and \ref{Tab:CosHeston} for increasing numbers of the discretization intervals $N_{\Delta t}$ and $N_{\Delta v}$. We can observe stability and convergence for both strategies but, looking at Table \ref{Tab:Heston_cpu}, the CPU-time saving of the COS BEM methodology is evident\footnote{The codes run on a laptop computer with Intel i5 CPU, 4Gb RAM. The algorithm is embarrassingly parallel; in particular the blocks $\mathcal{A}^{(\ell)}$ and $\mathcal{F}^{(\ell)}$ of the linear system are independent of each other, so the computation has been implemented with an OpenMP parallelization approach.}.\\\\
\begin{minipage}[h]{17.cm}\footnotesize
\begin{center}
$\begin{array}{cc}
\begin{minipage}[h]{5.cm}
\begin{tabular}{|c|c|}
  \multicolumn{2}{c}{BEM \cite{GuardasoniSanfelici2016}}\\   
  \hline
  $N_{\Delta t}=N_{\Delta v}$ & $V(150,0.01,0)$ \\
  \hline
  3 & 5.1021E+01\\
  6 & 5.1025E+01\\
  9 & 5.1024E+01\\
  12 & 5.1023E+01\\
  15 & 5.1023E+01\\
  \hline
\end{tabular}
 \end{minipage} &
\begin{minipage}[h]{5.cm}
\begin{tabular}{|c|c|}
  \multicolumn{2}{c}{COS BEM}\\   
  \hline
  $N_{\Delta t}=N_{\Delta v}$ & $V(150,0.01,0)$ \\
  \hline
  3 & 5.1021E+01\\
  6 & 5.1024E+01\\
  9 & 5.1023E+01\\
  12 & 5.1022E+01\\
  15 & 5.1022E+01\\
  \hline
\end{tabular}
 \end{minipage}
\end{array}$\end{center}\end{minipage}
\captionof{table}{Option values $V(150,0.01,0)$ computed by the BEM code (on the left) and by the COS BEM code (on the right) as a function of the discretization parameters $N_{\Delta t}$ and $N_{\Delta v}$.}
\label{Tab:Heston}\vspace{0.2cm}
\begin{minipage}[h]{17.cm}\footnotesize
\begin{center}
$\begin{array}{cc}
\begin{minipage}[h]{5.cm}
\begin{tabular}{|c|c|}
  \multicolumn{2}{c}{BEM \cite{GuardasoniSanfelici2016}}\\   
  \hline
  $N_{\Delta t}=N_{\Delta v}$ & $V(115,0.01,0)$ \\
  \hline
  3 & 8.3110E+00\\
  6 & 8.3244E+00\\
  9 & 8.3227E+00\\
  12 & 8.3220E+00\\
  15 & 8.3218E+00\\
  \hline
\end{tabular}
 \end{minipage} &
\begin{minipage}[h]{5.cm}
\begin{tabular}{|c|c|}
  \multicolumn{2}{c}{COS BEM}\\   
  \hline
  $N_{\Delta t}=N_{\Delta v}$ & $V(115,0.01,0)$ \\
  \hline
  3 & 8.3110E+00\\
  6 & 8.3204E+00\\
  9 & 8.3193E+00\\
  12 & 8.3190E+00\\
  15 & 8.3190E+00\\
  \hline
\end{tabular}
 \end{minipage}
\end{array}$\end{center}\end{minipage}
\captionof{table}{Option values $V(115,0.01,0)$ computed by the BEM code (on the left) and by the COS BEM code (on the right) as a function of the discretization parameters $N_{\Delta t}$ and $N_{\Delta v}$.}
\label{Tab:CosHeston}\vspace{0.2cm}

\begin{minipage}[h]{17.5cm}\footnotesize
\begin{center}
$\begin{array}{cc}
\begin{minipage}[h]{5.cm}
\begin{tabular}{|c|c|}
  \multicolumn{2}{c}{BEM \cite{GuardasoniSanfelici2016}}\\   
  \hline
  $N_{\Delta t}=N_{\Delta v}$ & CPU time \\
  \hline
  3 & 1.5E+02 s.\\
  6 & 7.5E+02 s.\\
  9 & 3.4E+03 s.\\
  12 & 3.7E+03 s.\\
  15 & 6.2E+03 s.\\
  \hline
\end{tabular}
 \end{minipage} &
\begin{minipage}[h]{5.cm}
\begin{tabular}{|c|c|}
  \multicolumn{2}{c}{COS BEM}\\
  \hline
  $N_{\Delta t}=N_{\Delta v}$ & CPU time \\
  \hline
  3 & 6.6E+00 s.\\
  6 & 1.7E+01 s.\\
  9 & 3.5E+01 s.\\
  12 & 6.5E+01 s.\\
  15 & 9.5E+01 s.\\
  \hline
\end{tabular}
 \end{minipage}
\end{array}$\end{center}\end{minipage}
\captionof{table}{Computation times of the BEM code (on the left) and the COS BEM code (on the right) as a function of the discretization parameters $N_{\Delta t}$ and $N_{\Delta v}$.}\label{Tab:Heston_cpu}\vspace{0.4cm}

\indent Moreover, we can observe that as $S$ gets closer to the barrier the computation becomes more challenging from the accuracy view point; nevertheless, the COS BEM method is less sensitive w.r.t. BEM to the tuning of quadrature parameters and in conclusion more reliable in the inverse transform.

\subsection{Approximation of an Up-and-Out Call Option}\label{sec:Heston_Up_OutCall}
Note that, the case of an up-and-out call option, can be easily recovered from the previous down-and-out case. In fact, in the face of a different asset domain definition for the differential problem:
\begin{eqnarray}
                \label{HestonPDE_bis}\displaystyle\frac{\partial u}{\partial t}+\frac{1}{2}v\frac{\partial^2 u}{\partial x^2}+\rho \eta v \frac{\partial^2 u}{\partial x \partial v}+\frac{1}{2}\eta^2 v
\frac{\partial^2 u}{\partial v^2} + \left(r-\delta-\frac{1}{2}v\right) \frac{\partial u}{\partial x}- (\lambda(v-\bar v)-\theta v) \frac{\partial u}{\partial v}=0, \hspace{1.cm}\\
                  \displaystyle x\in\Omega_x:=(-\infty,B)\,,\, v\in \Omega_v:=(0,+\infty)\,,\,t\in[0,T]\,\quad\nonumber\\\nonumber\\
u(x,v,T)=(e^{x}-E)^+=\max(e^x -E,0)\hspace{5.cm}x\in\Omega_x\,,\, v\in \Omega_v\quad\\
u(B,v,t)=0\hspace{8.9cm} v\in\Omega_v\,,\,t\in[0,T]\quad\label{BarrierCond_H_bis}
\end{eqnarray}
the representation formula \eqref{BatesForm1} and the BIE \eqref{BIE_H} are formally equal.\\ From the numerical point of view, as the required function $\frac{\partial u}{\partial y}$ is unknown at the same boundary half space $(B,w,\tau)$ with $w\in\Omega_v, t\in[0,T]$ of a down-and-out call option, the linear system has the same matrix entries \eqref{matrix_el_H} but different rhs
\begin{equation}\label{rhs_el_H_bis}
\begin{array}{rcl}
\mathcal{F}_i^{(j)} \!\!&=&\!\! \displaystyle\int_{-\infty}^{B}\max(e^y-E,0)\widetilde{G}(y,T;B,\overline{v}_i,\overline{t}_j)dy\vspace{0.2cm}\\
\!\!&=&\!\! \displaystyle\int_{\log(E)}^{B}(e^y-E)\widetilde{G}(y,T;B,\overline{v}_i,\overline{t}_j)dy\vspace{0.2cm}\\
\!\!&=&\!\! \displaystyle H[\min(b,B)-\max\big(a,\log(E)\big)]\int_{\max\big(a,\log(E)\big)}^{\min(b,B)}\!\!\!\!\!\!\!(e^y-E)\sum_{n=0}^{N_F-1}{}^{\prime}F_n[B,\overline{v}_i,T,\overline{t}_j]\cos\left(n\pi\frac{y-a}{b-a}\right) dy\vspace{0.2cm}\\
\!\!&=&\!\! \displaystyle H[\min(b,B)-\max\big(a,\log(E)\big)]
\sum_{n=0}^{N_F-1}{}^{\prime}F_n[B,\overline{v}_i,T,\overline{t}_j]V_n
\end{array}
\end{equation}
with
\begin{equation}\label{F_k_H_rhs_bis}
\begin{array}{l}
\displaystyle F_n[x,v,T,t]=\frac{2}{b-a}\textrm{Re}\left\{\phi_T\left(\frac{n\pi}{b-a};x,v,t\right)e^{-\mathbf{i}n\pi\frac{a}{b-a}}\right\},\qquad n=0,\ldots,N_F-1,\vspace{0.2cm}\\
\displaystyle V_n=\int_{\max\big(a,\log(E)\big)}^{\min(b,B)}\!\!\!\!\!(e^y-E)\cos\left(n\pi\frac{y-a}{b-a}\right)dy, \qquad n=0,\ldots,N_F-1,\quad \textrm{analitically integrated}
\end{array}
\end{equation}
with bounds $a,b$ as in \eqref{bounds_H_rhs}.\\
Again, in the post-processing, for the evaluation of the undiscounted price of barrier option $u(x,v,t)$ in $\Omega_x\times\Omega_v$ for $t\in[0,T)$, the equation (\ref{BatesForm1}) is modified only in the first term on the right-hand side:
\begin{equation}\label{postpro_H_bis}
\begin{array}{l}
\displaystyle u(x,v,t)\approx \int^{B}_{-\infty}\max(e^y-E,0)\widetilde{G}(y,T;x,v,t)dy+\vspace{0.2cm}\\
\displaystyle -\sum_{h=1}^{N_{\Delta v}}\sum_{k=\textrm{floor}[\frac{t}{\Delta t}]+1}^{N_{\Delta t}}\alpha_h^{(k)}\int_{\max(t,t_{k-1})}^{t_k}\int_{v_{h-1}}^{v_h}\frac{w}{2}G(B,w,\tau;x,v,t)dw\,d\tau\,.
\end{array}
\end{equation}
manipulated as the elements ${\cal F}_i^{(j)}$ of equations \eqref{rhs_el_H_bis}-\eqref{F_k_H_rhs_bis}):
\begin{equation}\label{postpro_initdata_H_bis}
\begin{array}{l}
    \displaystyle \int^{B}_{-\infty}\max(e^y-E,0)\widetilde{G}(y,T;x,v,t)dy=H[\min(b,B)-\max\big(a,\log(E)\big)]
\sum_{n=0}^{N_F-1}{}^{\prime}F_n[x,v,T,t]V_n
\end{array}
\end{equation}
with bounds $[a,b]$ as in \eqref{bounds_H_rhs}.\\
Some numerical results for comparison are available in \cite{Chiarella2012}, setting the data:
$$
\begin{array}{|c|c|c|c|c|c|c|c|c|}
  \hline
  \lambda & \bar{v} & \rho & \eta & r & \delta & E & e^B & T \\
  \hline
  2 & 0.1 & -0.5 & 0.1 & 0.03 & 0.05 & 100 & 130 & 0.5\\
  \hline
\end{array}\vspace{-0.2cm}
$$\captionof{table}{Heston model. Up-and-out call option data. The stochastic volatility parameters are those used in Heston's original paper \cite{Heston1993}.}\label{Tab:H_data_bis}\vspace{0.2cm}
The present option prices are displayed in Table \ref{Tab:Heston_bis_1} for several asset values and at the current variance value $v=0.1$ for increasing numbers of the discretization intervals $N_{\Delta t}=N_{\Delta v}$. We can observe stability, convergence and good agreement with results in \cite{Chiarella2012} (look at Table \ref{Tab:Chiarella1}).\\
$$
\begin{array}{clllll}
  \hline
	V(S,0.1,0) & S\\\cline{2-6}
  N_{\Delta t}=N_{\Delta v} & 80 & 90 & 100 & 110 & 120\\
  \hline
  6 & 0.9113\quad & 1.8855\quad & 2.5858\quad & 2.4474\quad & 1.4255\quad\\
	9 & 0.9082 & 1.8823 & 2.5908 & 2.4713 & 1.4738\\
	12 & 0.9074 & 1.8793 & 2.5904 & 2.4722 & 1.4704\\
  \hline
\end{array}\vspace{-0.2cm}
$$\captionof{table}{Heston model. Up-and-out call option data with data in Table \ref{Tab:H_data_bis}.}\label{Tab:Heston_bis_1}\vspace{0.2cm}
\begin{center}
	\includegraphics[scale=0.11]{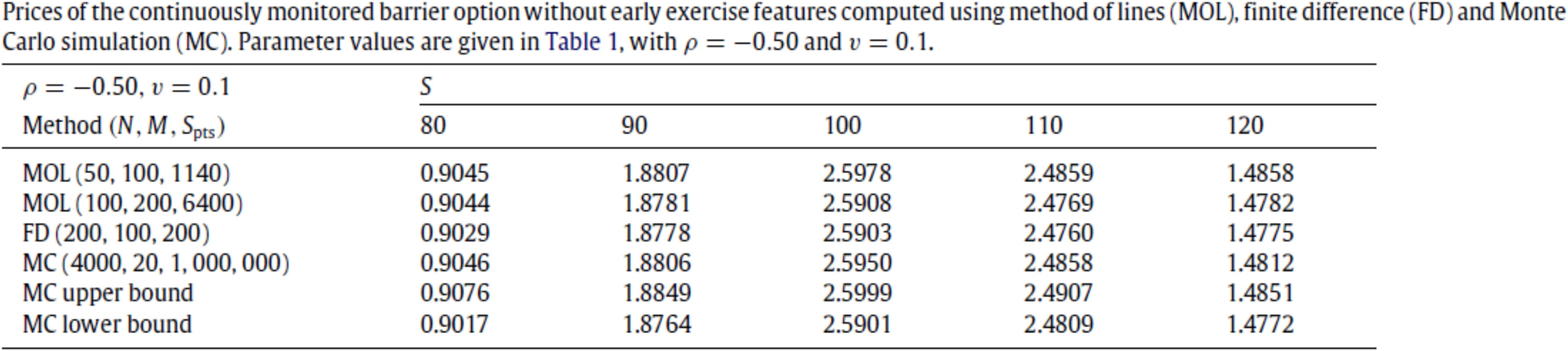}
	\caption{Numerical results in \cite{Chiarella2012}.}\label{Tab:Chiarella1}
\end{center}

The numerical approximation of $\Delta-$greek in \eqref{Delta} reduces to the approximation of $\frac{\partial u}{\partial x}(x,v,t)$ as $\frac{\partial V}{\partial S}(S,v,t)=e^{-x}\frac{\partial u}{\partial x}(x,v,t)$. With our method this can be done without computing the option value $V$, directly in the post-processing phase, observing that
\begin{equation}\label{postpro_H_Delta}
\begin{array}{l}
\displaystyle \frac{\partial u}{\partial x}(x,v,t)\, \approx \int^{B}_{-\infty}\max(e^y-E,0)\frac{\partial \widetilde{G}}{\partial x}(y,T;x,v,t)dy+\vspace{0.2cm}\\
\displaystyle -\sum_{h=1}^{N_{\Delta v}}\sum_{k=\textrm{floor}[\frac{t}{\Delta t}]+1}^{N_{\Delta t}}\alpha_h^{(k)}\int_{\max(t,t_{k-1})}^{t_k}\int_{v_{h-1}}^{v_h}\frac{w}{2}\frac{\partial G}{\partial x}(B,w,\tau;x,v,t)dw\,d\tau\,.
\end{array}
\end{equation}
The first term on the right-hand side of (\ref{postpro_H_Delta}) is:
\begin{equation}
\begin{array}{l}
    \displaystyle \int^{B}_{-\infty}\max(e^y-E,0)\frac{\partial \widetilde{G}}{\partial x}(y,T;x,v,t)dy=H[\min\big(b,B\big)-\max\big(a,\log(E)\big)]
\sum_{n=0}^{N_F-1}{}^{\prime} F_n[0,v,T,t]V_n
\end{array}
\end{equation}
with 
\begin{equation}
\begin{array}{l}
\displaystyle F_n[0,v,T,t]=\frac{2}{b-a}\textrm{Re}\left\{-\phi_T\left(\frac{n\pi}{b-a};0,v,t\right)e^{-\mathbf{i}n\pi\frac{a}{b-a}}\right\},\quad n=0,\ldots,N_F-1,\vspace{0.2cm}\\
\displaystyle V_n=\int_{\max\big(a,\log(E)\big)}^{\min(b,B)}\!\!\!\!\!(e^y-E)\sin\left(n\pi\frac{y-x-a}{b-a}\right)\frac{n\pi}{b-a}dy, \quad n=0,\ldots,N_F-1,\quad \textrm{analytically integrated}
\end{array}
\end{equation}
and bounds $[a,b]$ as in \eqref{bounds_H_rhs}.\\
The second term on the right-hand side of (\ref{postpro_H_Delta}) evaluated at $t=0$ is
\begin{equation}
\begin{array}{l}
    \displaystyle -\sum_{h=1}^{N_{\Delta v}}\sum_{k=1}^{N_{\Delta t}} \alpha_h^{(k)}\int_{t_{k-1}}^{t_k}\int_{v_{h-1}}^{v_h}\frac{w}{2}\frac{\partial G}{\partial x}(B,w,\tau;x,v,0)dw\,d\tau=\vspace{0.2cm}\\
\displaystyle=-\sum_{h=1}^{N_{\Delta v}}\sum_{k=1}^{N_{\Delta t}} \alpha_h^{(k)}\int_{t_{k-1}}^{t_k}\int_{v_{h-1}}^{v_{h}}
\frac{w}{2}p_v(w,\tau|v)\frac{\partial p}{\partial x}(B-x,\tau|w,v)dw\,d\tau=\vspace{0.2cm}\\
\displaystyle=-\sum_{h=1}^{N_{\Delta v}}\sum_{k=1}^{N_{\Delta t}} \alpha_h^{(k)}\int_{0}^{1}\int_{v_{h-1}}^{v_{h}}
\!\!\frac{\Delta t}{2}w\,p_{v}(w,\Delta t(k-1+s)|v)\vspace{0.2cm}\\
\displaystyle\quad\sum_{n=0}^{N_F-1}{}^{\prime}F_n[v,w,\Delta t(k-1+s)]\sin\left(n\pi\frac{B-x-a}{b-a}\right)\frac{n\pi}{b-a} dw\, ds
\end{array}
\end{equation}
and it can be computed as done for the linear system entries by \eqref{F_k_H_matrix} and \eqref{bounds_H_matrix}.\\
In Fig. \ref{Fig:Delta}, the Delta profile obtained with the parameters in Table \ref{Tab:H_data_bis} is shown for some values of current variances, in good agreement with the results in the paper \cite{Chiarella2012}.
\begin{center}
	\includegraphics[trim=500 50 500 50, clip,scale=0.6]{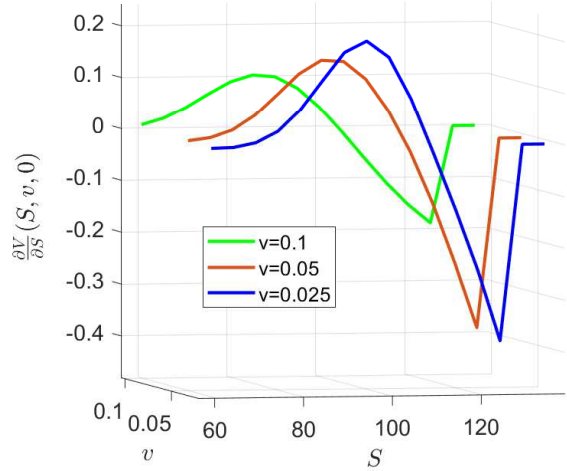}
	\captionof{figure}{Delta profile for a European up-and-out call option.}\label{Fig:Delta}
\end{center}

\subsection{Heston model for an Up-and-Out Cash-or-Nothing Call Option}\label{sec:Heston_Cash_Noth}
Changing the payoff implies changing the known terms in the equations. An advantage of the BEM approach is that the irregularities in the payoff function do not affect the stability and the accuracy of the option evaluation, since this numerical method relies on an integral formulation and not on finite differences that may magnify discontinuities.\\ Starting from the model problem, the changes for a cash-or-nothing payoff function are:
\begin{eqnarray}
                \label{HestonPDE_tris}\displaystyle\frac{\partial u}{\partial t}+\frac{1}{2}v\frac{\partial^2 u}{\partial x^2}+\rho \eta v \frac{\partial^2 u}{\partial x \partial v}+\frac{1}{2}\eta^2 v
\frac{\partial^2 u}{\partial v^2} + \left(r-\delta-\frac{1}{2}v\right) \frac{\partial u}{\partial x}- (\lambda(v-\bar v)-\theta v) \frac{\partial u}{\partial v}=0, \hspace{1.cm}\\
                  \displaystyle x\in\Omega_x:=(-\infty,B)\,,\, v\in \Omega_v:=(0,+\infty)\,,\,t\in[0,T]\,\quad\nonumber\\\nonumber\\
u(x,v,T)=H[e^{x}-E]\hspace{7.8cm}x\in\Omega_x\,,\, v\in \Omega_v\quad\\
u(B,v,t)=0\hspace{8.9cm} v\in\Omega_v\,,\,t\in[0,T]\quad\label{BarrierCond_H_tris}
\end{eqnarray}
whose related representation formula is 
\begin{equation}\label{BatesForm1_tris}
\begin{array}{r}
\displaystyle u(x,v,t)=\int_{\Omega_x}\int_{\Omega_v} H[e^y-E] G(y,w,T;x,v,t)dw\,dy-\!\int_t^T\!\! \int_{\Omega_v}  \frac{w}{2}G(B,w,s;x,v,t) \frac{\partial u}{\partial y}(B,w,s) dw\, ds\,\vspace{0.2cm}\\\displaystyle
x\in\Omega_x\,,\,v\in \Omega_v\,,\,t\in[0,T]
\end{array}
\end{equation}
and the BIE
\begin{equation}\label{BIE_H_tris}
\begin{array}{l}
\displaystyle 0=u(B,v,t)= \!\! \int_{\Omega_x}\int_{\Omega_v}\!\!H[e^y-E]G(y,w,T;B,v,t)dw\,dy+\vspace{0.2cm}\\
\displaystyle -\int_{t}^T\int_{\Omega_v}\frac{w}{2}G(B,w,\tau;B,v,t)\frac{\partial u}{\partial y}(B,w,\tau)dw\,d\tau\,.
\end{array}
\end{equation}
From the numerical point of view, the payoff modification affects only the right-hand side entries
\begin{equation}\label{rhs_el_H_tris}
\begin{array}{rcl}
\mathcal{F}_i^{(j)} \!\!&=&\!\! \displaystyle\int_{-\infty}^{B}H[e^y-E]\widetilde{G}(y,T;B,\overline{v}_i,\overline{t}_j)dy=\int_{\log(E)}^{B}\widetilde{G}(y,T;B,\overline{v}_i,\overline{t}_j)dy\vspace{0.2cm}\\
\!\!&=&\!\! \displaystyle H[\min(b,B)-\max\big(a,\log(E)\big)]\int_{\max\big(a,\log(E)\big)}^{\min(b,B)}\!\!\sum_{n=0}^{N_F-1}{}^{\prime}F_n[B,\overline{v}_i,T,\overline{t}_j]\cos\left(n\pi\frac{y-a}{b-a}\right) dy\vspace{0.2cm}\\
\!\!&=&\!\! \displaystyle H[\min(b,B)-\max\big(a,\log(E)\big)]
\sum_{n=0}^{N_F-1}{}^{\prime}F_n[B,\overline{v}_i,T,\overline{t}_j]V_n
\end{array}
\end{equation}
with
\begin{equation}\label{F_k_H_rhs_tris}
\begin{array}{l}
\displaystyle F_n[x,v,T,t]=\frac{2}{b-a}\textrm{Re}\left\{\phi_T\left(\frac{n\pi}{b-a};x,v,t\right)e^{-\mathbf{i}n\pi\frac{a}{b-a}}\right\},\qquad n=0,\ldots,N_F-1,\vspace{0.2cm}\\
\displaystyle V_n=\int_{\max\big(a,\log(E)\big)}^{\min(b,B)}\!\!\cos\left(n\pi\frac{y-a}{b-a}\right)dy, \qquad n=0,\ldots,N_F-1,\quad \textrm{analytically integrated}
\end{array}
\end{equation}
and the first term of the post-processing:
\begin{equation}\label{postpro_H_tris}
\begin{array}{l}
\displaystyle u(x,v,t)\approx \int^{B}_{-\infty}H[e^y-E]\widetilde{G}(y,T;x,v,t)dy+\vspace{0.2cm}\\
\displaystyle -\sum_{h=1}^{N_{\Delta v}}\sum_{k=\textrm{floor}[\frac{t}{\Delta t}]+1}^{N_{\Delta t}}\alpha_h^{(k)}\int_{\max(t,t_{k-1})}^{t_k}\int_{v_{h-1}}^{v_h}\frac{w}{2}G(B,w,\tau;x,v,t)dw\,d\tau\,.
\end{array}
\end{equation}
After some manipulations as for the elements ${\cal F}_i^{(j)}$ of equations \eqref{rhs_el_H_tris}-\eqref{F_k_H_rhs_tris}), we get:
\begin{equation}\label{postpro_initdata_H_tris}
\begin{array}{l}
    \displaystyle \int^{B}_{-\infty}H[e^y-E]\widetilde{G}(y,T;x,v,t)dy=H[\min(b,B)-\max\big(a,\log(E)\big)]
\sum_{n=0}^{N_F-1}{}^{\prime}F_n[x,v,T,t]V_n\,.
\end{array}
\end{equation}
For both, the bounds $a,b$ are defined as in \eqref{bounds_H_rhs}.\\
With the parameters in Table \ref{Tab:H_data_CashNoth}
$$
\begin{array}{|c|c|c|c|c|c|c|c|c|}
  \hline
  \lambda & \bar{v} & \rho & \eta & r & \delta & E & e^B & T \\
  \hline
  4 & 0.04 & \pm 0.5 & 0.1 & 0.05 & 0.02 & 100 & 110 & 1\\
  \hline
\end{array}
$$\captionof{table}{Heston model. Cash-or-nothing up-and-out option data.}\label{Tab:H_data_CashNoth}\vspace{0.2cm}
the option value without barriers is that shown in Fig. \ref{Fig:CashNoth} and, in Fig. \ref{Fig:CashNoth_barrier}, we observe that the introduction of a barrier changes the shape and reduces the option value as expected.
\begin{center}
	\includegraphics[scale=0.5]{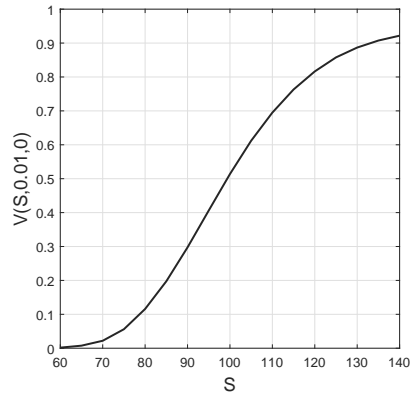}
	\captionof{figure}{European cash-or-nothing option profile for $\rho=-0.5$ and $v=0.01$.}\label{Fig:CashNoth}
\end{center}
\begin{center}
	\includegraphics[trim=450 0 500 0, clip,scale=0.44]{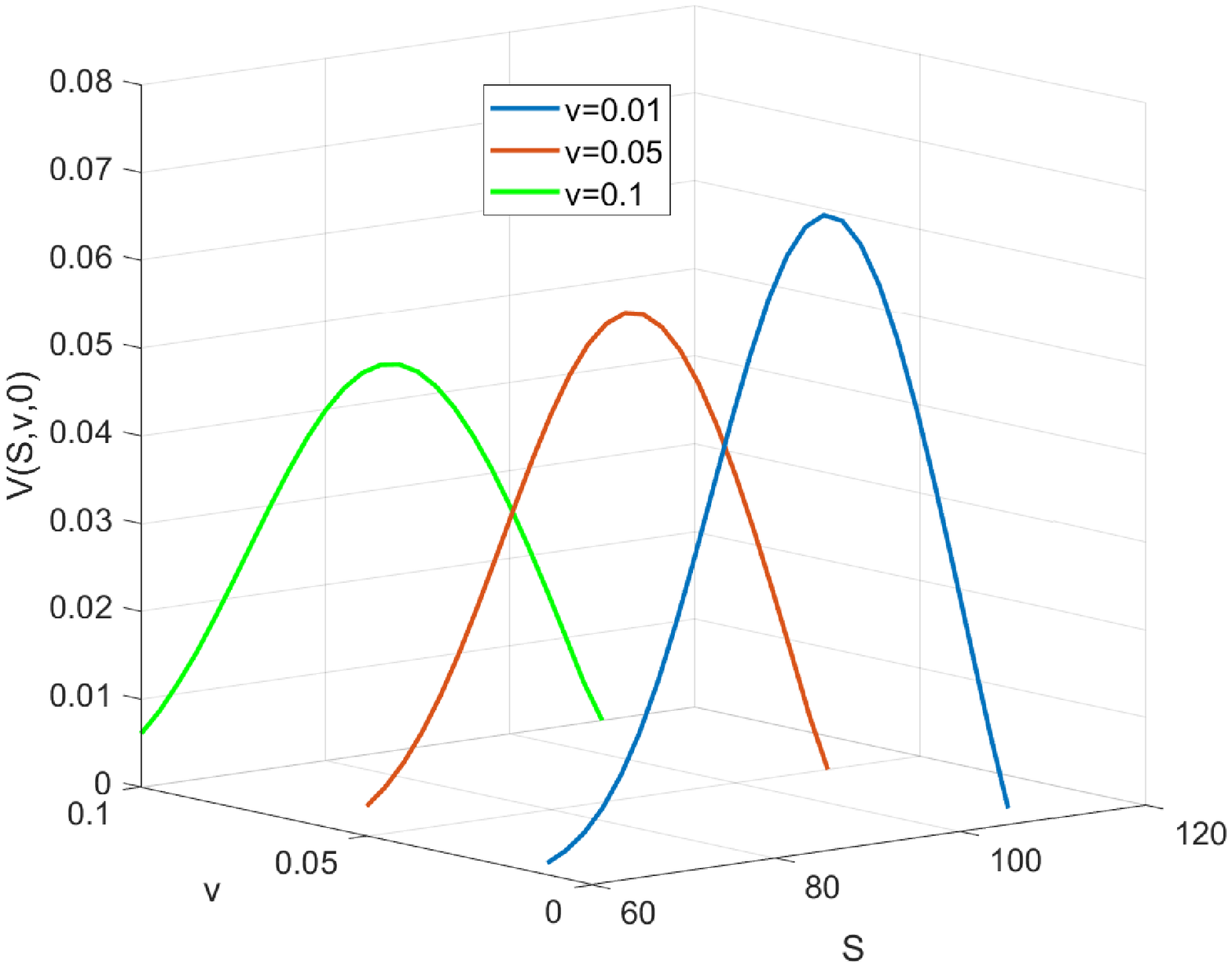}\,
	\includegraphics[trim=440 0 500 0, clip,scale=0.44]{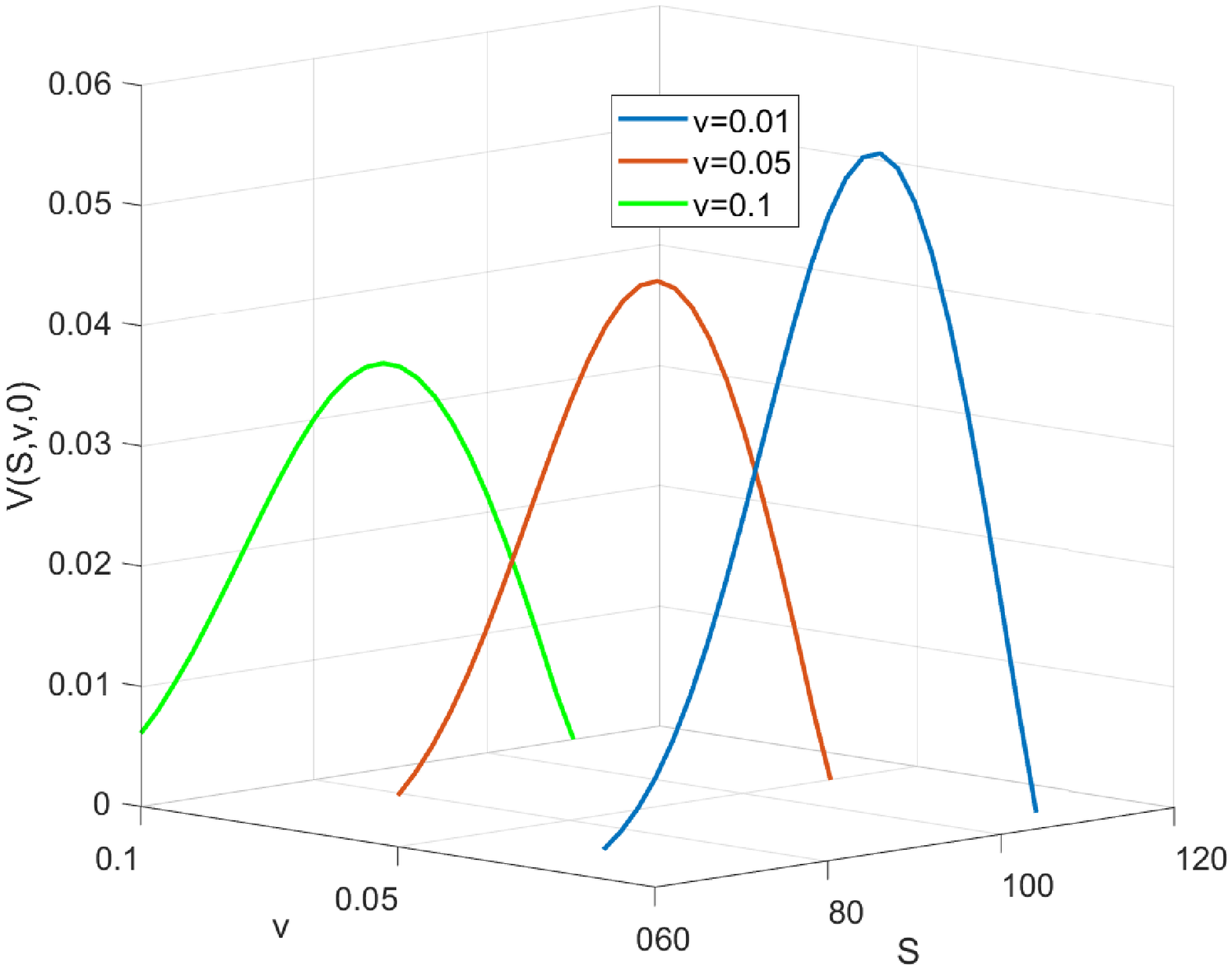}
	\captionof{figure}{European cash-or-nothing up-and-out option profile for $\rho=-0.5$ (on the left) and $\rho=0.5$ (on the right) and some current variances $v$.}
	\label{Fig:CashNoth_barrier}
\end{center}
In order to validate COS BEM, we illustrate a comparison with the conditional MC method \cite{Glasserman2004}, denoting by $N$ the number of random samples and by $M$ the number of time steps in Euler time discretization as in \cite{GuardasoniSanfelici2016}. As COS BEM, also the MC code has been parallelized to exploit the Intel i5 CPU features. In Figure \ref{Fig: BEM_MC}, we plot the graphs obtained for the current variance $v=0.01$, by COS BEM ($N_{\Delta t}=N_{\Delta v}=6$) and by MC ($M=100$ and $N=10^4$) with the related confidence interval. The COS BEM curve lies within the confidence interval, but the variability of the MC estimation is evident.
\begin{center}
	\includegraphics[scale=0.6]{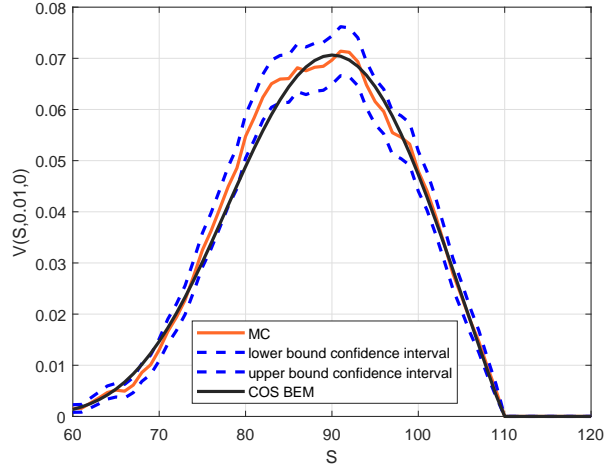}
	\captionof{figure}{European cash-or-nothing option: comparison between MC and COS BEM.}\label{Fig: BEM_MC}
\end{center}
Then we observe that, if we need only a very poor approximation of the option value (with an accuracy of approximately 1 decimal digit) we may be satisfied by a MC estimate obtainable with $N=10^6$ in less than 1 second. Nevertheless, if we require a greater accuracy then COS BEM overcomes MC and shows a clear trend of convergence without exceeding 10 seconds of CPU time (see Tables \ref{Tab:S100_BEM}-\ref{Tab:S100_MC_time}). On the contrary, Monte Carlo appears to require a number of samples $N$ greater than $10^8$ to get an acceptable accuracy, but using $N=10^{10}$ requires a CPU time one order of magnitude ($\mathcal{O}(10^1) s.$) larger. We remark that, as expected, the increase of $M$ has no positive effects on the accuracy that seems to be dominated by the sampling error (see Tables \ref{Tab:S100_MC}-\ref{Tab:S109_MC}). Instead, it may cause a shift of the confidence intervals, so that they may happen to be disjoint: looking at Table \ref{Tab:S100_MC}, the confidence interval obtained with $N=10^6$ and $M=800$ is disjoint with respect to all the confidence intervals of the same column.\\ Observing in deeper the numerical results of COS BEM in Table \ref{Tab:S100_BEM}, we observe that rising $N_v$ from 25 to 30 is not so significant as instead refining the time discretization.\\ Moreover the closer the asset is to the barrier (Tables \ref{Tab:S109_BEM} and \ref{Tab:S109_MC}), the more we have to refine the approximation, requiring a greater computational effort and this applies for both the methods. \\
\begin{minipage}[h]{16.cm}
\begin{center}
\begin{tabular}{|c|c|c|c|c|c|}
  \multicolumn{6}{c}{COS BEM: $S=100$}\\
  \hline
  $N_{\Delta t}$ & $N_{\Delta v}=$10 & $N_{\Delta v}=$15 & $N_{\Delta v}=$20 & $N_{\Delta v}=$25 & $N_{\Delta v}=$30\\
  \hline
  40 & 4.7858E-02 & 4.7831E-02 & 4.7824E-02 & 4.7821E-02 & 4.7820E-02\\
  \hline
  60 & 4.7876E-02 & 4.7850E-02 & 4.7843E-02 & 4.7840E-02 & 4.7839E-02\\
  \hline
  80 & 4.7884E-02 & 4.7858E-02 & 4.7852E-02 & 4.7849E-02 & 4.7847E-02\\
  \hline
  100 & 4.7889E-02 & 4.7863E-02 & 4.7856E-02 & 4.7854E-02 & 4.7852E-02\\
  \hline
\end{tabular}\vspace{-0.0cm}
\end{center}\end{minipage}\captionof{table}{Option value approximations of $V(100,0.01,0)$ obtained by COS BEM, with respect to the number of discretization intervals.}\label{Tab:S100_BEM}
\begin{minipage}[h]{16.cm}
\begin{center}
\begin{tabular}{|c|c|c|c|c|c|c|c|}
  \multicolumn{8}{c}{COS BEM: $S=109$}\\
  \hline
  $N_{\Delta t}$ & $N_{\Delta v}=$10 & $N_{\Delta v}=$15 & $N_{\Delta v}=$20 & $N_{\Delta v}=$25 & $N_{\Delta v}=$30 & $N_{\Delta v}=$35  & $N_{\Delta v}=$40\\
  \hline
  40 & 4.5724E-03 & 4.5669E-03 & 4.5615E-03 & 4.5611E-03 & 4.5598E-03 & 4.5593E-03 & 4.5590E-03\\
  \hline
  80 & 4.5848E-03 & 4.5793E-03 & 4.5740E-03 & 4.5736E-03 & 4.5724E-03 & 4.5720E-03 & 4.5717E-03\\
  \hline
  120 & 4.5876E-03 & 4.5821E-03 & 4.5768E-03 & 4.5765E-03 & 4.5754E-03 & 4.5749E-03 & 4.5747E-03\\
  \hline
  160 & 4.5888E-03 & 4.5833E-03 & 4.5780E-03 & 4.5777E-03 & 4.5766E-03 & 4.5762E-03 & 4.5760E-03\\
  \hline
  200 & 4.5895E-03 & 4.5840E-03 & 4.5786E-03 & 4.5784E-03 & 4.5773E-03 & 4.5769E-03 & 4.5767E-03\\
  \hline 
  240 & 4.5899E-03 & 4.5844E-03 & 4.5791E-03 & 4.5788E-03 & 4.5778E-03 & 4.5774E-03 & 4.5772E-03\\
  \hline
\end{tabular}\vspace{-0.0cm}
\end{center}\end{minipage}\captionof{table}{Option value approximations of $V(109,0.01,0)$ obtained by COS BEM, with respect to the number of discretization intervals.}\label{Tab:S109_BEM}
\begin{minipage}[h]{16.cm}
\begin{center}
\begin{tabular}{|c|c|c|c|c|c|}
  \multicolumn{6}{c}{COS BEM CPU-time: $S=100$}\\
  \hline
  $N_{\Delta t}$ & $N_{\Delta v}=$10 & $N_{\Delta v}=$15 & $N_{\Delta v}=$20 & $N_{\Delta v}=$25 & $N_{\Delta v}=$30\\
  \hline
  40 & 2.8E+00 & 3.0E+00 & 3.6E+00 & 5.6E+00 & 6.3E+00\\
  \hline
  60 & 3.3E+00 & 2.9E+00 & 3.9E+00 & 6.0E+00 & 6.8E+00\\
  \hline
  80 & 5.8E+00 & 5.4E+00 & 6.4E+00 & 8.2E+00 & 7.9E+00\\
  \hline
  100 & 4.3E+00 & 6.2E+00 & 5.9E+00 & 7.8E+00 & 9.8E+00\\
  \hline
\end{tabular}\vspace{-0.0cm}
\end{center}\end{minipage}\captionof{table}{CPU-time of COS BEM method for the evaluation of $V(100,0.01,0)$.}\label{Tab:S100_BEM_time}
\begin{minipage}[h]{16.cm}
\begin{center}\small
\begin{tabular}{|r|c|c|c|c|c|c|}
\multicolumn{7}{c}{MC, S=100} \\
\hline
\multicolumn{1}{|c|}{M}  & $N=10^4$ &\!\!\!\! \mbox{95\% conf. int.}\!\!& $N=10^6$ &\!\!\!\! 95\% conf. int.\!\! & $N=10^8$ &\!\!\!\! 95\% conf. int.\!\! \\
    \hline
        200 & 4.8239E-02 & [4.42E-02,5.22E-02] & 4.7941E-02 & [4.75E-02,4.83E-02] & 4.7805E-02 & [4.776E-02,4.784E-02]\\ 
        400 & 4.6817E-02 & [4.29E-02,5.08E-02] & 4.8155E-02 & [4.78E-02,4.86E-02] & 4.7836E-02 & [4.780E-02,4.788E-02]\\ 
        800 & 4.2510E-02 & [3.87E-02,4.63E-02] & 4.8628E-02 & [4.82E-02,4.90E-02] & 4.7854E-02 & [4.781E-02,4.789E-02]\\
        1600 & 5.0459E-02 & [4.63E-02,5.46E-02] & 4.7895E-02 & [4.75E-02,4.83E-02] & 4.7827E-02 & [4.779E-02,4.787E-02]\\
        3200 & 5.1261E-02 & [4.71E-02,5.54E-02] & 4.7787E-02 & [4.74E-02,4.82E-02] & 4.7863E-02 & [4.782E-02,4.790E-02]\\
        \hline
\end{tabular}
\end{center}\end{minipage}\captionof{table}{Option value approximations of $V(100,0.01,0)$ and 95\% confidence intervals obtained using Monte Carlo method, with respect to the number of samples $N$ and the number of time steps $M$.}\label{Tab:S100_MC}\vspace{0.2cm}
\begin{minipage}[h]{16.cm}
\begin{center}\small
\begin{tabular}{|r|c|c|c|c|c|c|}
\multicolumn{7}{c}{MC, S=109} \\
\hline
\multicolumn{1}{|c|}{M}  & $N=10^4$ &\!\!\!\! \mbox{95\% conf. int.}\!\!& $N=10^6$ &\!\!\!\! 95\% conf. int.\!\! & $N=10^8$ &\!\!\!\! 95\% conf. int.\!\! \\
    \hline
        200 & 4.7216E-03 & [3.53E-03,5.91E-03] & 4.5860E-03 & [4.47E-03,4.70E-03] & 4.5624E-03 & [4.55E-03,4.57E-03]\\
        400 & 3.8892E-03 & [2.78E-03,5.00E-03] & 4.5605E-03 & [4.44E-03,4.68E-03] & 4.5825E-03 & [4.57E-03,4.59E-03]\\
        800 & 4.8720E-03 & [3.58E-03,6.16E-03] & 4.6214E-03 & [4.50E-03,4.75E-03] & 4.5889E-03 & [4.58E-03,4.60E-03]\\
        1600 & 4.3290E-03 & [3.13E-03,5.53E-03] & 4.5306E-03 & [4.41E-03,4.66E-03] & 4.5803E-03 & [4.57E-03,4.59E-03]\\
        3200 & 3.9856E-03 & [2.81E-03,5.16E-03] & 4.6252E-03 & [4.50E-03,4.75E-03] & 4.5845E-03 & [4.57E-03,4.60E-03]\\\hline
\end{tabular}
\end{center}\end{minipage}\captionof{table}{Option value approximations of $V(109,0.01,0)$ and 95\% confidence intervals obtained using Monte Carlo method, with respect to the number of samples $N$ and the number of time steps $M$.}\label{Tab:S109_MC}\vspace{0.2cm}
\begin{minipage}[h]{16.cm}
\begin{center}
\begin{tabular}{|r|l|l|l|}
\multicolumn{4}{c}{Monte Carlo CPU-time}\\
\hline
\multicolumn{1}{|c|}{$M$}  & $N=10^4$ & $N=10^6$ & $N=10^8$ \\
\hline
200  & 1.7E-01 s. & 4.1E-01 s. & 7.7E+00 s. \\
400  & 4.0E-01 s. & 1.6E-01 s. & 1.6E+00 s. \\
800  & 4.4E-01 s. & 5.5E-01 s. & 2.6E+00 s. \\
1600 & 2.3E-01 s. & 1.5E+00 s. & 3.2E+00 s.\\
3200 & 3.8E-01 s. & 1.5E+00 s. & 9.3E+00 s.\\
\hline
\end{tabular}
\end{center}\end{minipage}
\captionof{table}{CPU-time of Monte Carlo method for the evaluation of $V(100,0.01,0)$.}\label{Tab:S100_MC_time}

\section{Conclusions and future work}
\label{sec;Conclusions}
In this work, the COS method \cite{FangOosterlee2008} has been combined with the BEM introduced in \cite{GuardasoniSanfelici2016} for a faster evaluation of barrier option prices in the Heston model framework. Obviously, this methodology can be adapted and extended to other asset models where the PDF is not explicitly known. In fact, the COS method has proved to be an efficient and stable technique for computing the numerical Fourier inverse transform of the characteristic function evaluated at one point only, overcoming defects of quadrature and other methods with the same purpose. 
Despite the efficiency of the COS BEM method, there are two aspects that might hamper its practical use. The first issue is the determination of the truncation interval, recently tackled by \cite{Junike2022}, while the second is the non a priori knowledge of the number of terms used in the cosines expansion. We have addressed this last challenging issue with an error analysis based on the characteristic function of the log-asset price.\\

\textbf{Funding}: This work has been partially supported by INdAM-GNCS Research Projects as well as by grants PID2019-105986GB-C21 and PID2020-118339GB-I00 from the Spanish Ministry of Economy and Competitiveness, and grant 2020-PANDE-00074 from the Secretaria d'Universitats i Recerca del departament d'Empresa i Coneixement de la Generalitat de Catalunya.

\bibliographystyle{alpha}
\bibliography{main}


\end{document}